\documentclass[aps,twocolumn,superscriptaddress,floatfix,nofootinbib,showpacs,altaffilletter]{revtex4-2}

\usepackage{epsfig,graphicx,amssymb,amsmath,wasysym,graphicx}

\usepackage{amsmath}
\usepackage{amsfonts}
\usepackage{amssymb}
\usepackage{graphicx}
\usepackage{bm}
\usepackage{subfigure}
\usepackage{url}
\usepackage[hyperindex]{hyperref}
\usepackage{color}
\usepackage[ddmmyy,24hr]{datetime}
\usepackage{bigdelim}
\usepackage{booktabs}
\usepackage{dcolumn}
\usepackage{multirow}
\usepackage{subfigure}
\usepackage{cancel}
\usepackage{stackrel}
\usepackage{paralist}
\usepackage{xspace}
\usepackage{slashed}
\usepackage{cancel}
\usepackage{todonotes}
\usepackage{enumerate}
\usepackage{float}
\usepackage{fullpage}
\usepackage{ulem}
\usepackage{physics}

\usepackage{lineno}

\newcommand{\nua}[1]{\ensuremath{\rlap{\kern-2.5pt\ensuremath{\overset{\scriptscriptstyle(-)}{\phantom{\nu}}}}{\ensuremath{{\nu}_{#1}}}}}

\usepackage{enumitem}

\newcommand{\cenns}{CE$\nu$NS\xspace}

\newcommand{\be}{\begin{equation}}
\newcommand{\ee}{\end{equation}}
\newcommand{\ba}{\begin{array}}
\newcommand{\ea}{\end{array}}

\newcommand{\wma}{{$\sin^2\!\vartheta_{\text{W}}\,$}}

\begin{document}

\title{Refined determination of the weak mixing angle at low energy}

\author{M. Atzori Corona}
\email{mattia.atzori.corona@ca.infn.it}
\affiliation{Dipartimento di Fisica, Universit\`{a} degli Studi di Cagliari,
	Complesso Universitario di Monserrato - S.P. per Sestu Km 0.700,
	09042 Monserrato (Cagliari), Italy}
\affiliation{Istituto Nazionale di Fisica Nucleare (INFN), Sezione di Cagliari,
	Complesso Universitario di Monserrato - S.P. per Sestu Km 0.700,
	09042 Monserrato (Cagliari), Italy}

\author{M. Cadeddu}
\email{matteo.cadeddu@ca.infn.it}
\affiliation{Istituto Nazionale di Fisica Nucleare (INFN), Sezione di Cagliari,
	Complesso Universitario di Monserrato - S.P. per Sestu Km 0.700,
	09042 Monserrato (Cagliari), Italy}

\author{N. Cargioli}
\email{nicola.cargioli@ca.infn.it}
\affiliation{Dipartimento di Fisica, Universit\`{a} degli Studi di Cagliari,
	Complesso Universitario di Monserrato - S.P. per Sestu Km 0.700,
	09042 Monserrato (Cagliari), Italy}
\affiliation{Istituto Nazionale di Fisica Nucleare (INFN), Sezione di Cagliari,
	Complesso Universitario di Monserrato - S.P. per Sestu Km 0.700,
	09042 Monserrato (Cagliari), Italy}

\author{F. Dordei}
\email{francesca.dordei@cern.ch}
\affiliation{Istituto Nazionale di Fisica Nucleare (INFN), Sezione di Cagliari,
	Complesso Universitario di Monserrato - S.P. per Sestu Km 0.700,
	09042 Monserrato (Cagliari), Italy}

\author{C. Giunti}
\email{carlo.giunti@to.infn.it}
\affiliation{Istituto Nazionale di Fisica Nucleare (INFN), Sezione di Torino, Via P. Giuria 1, I--10125 Torino, Italy}

\begin{abstract}

The weak mixing angle is a fundamental parameter of the electroweak theory of the standard model whose measurement in the low-energy regime is still not precisely determined. Different probes are sensitive to its value, among which atomic parity violation, coherent elastic neutrino-nucleus scattering and parity-violating electron scattering on different nuclei. In this work, we attempt for the first time to combine all these various determinations by performing a global fit that also keeps into account the unavoidable dependence on the experimentally poorly known neutron distribution radius of the nuclei employed, for which a new measurement using proton-cesium elastic scattering became available. By using all present direct determinations of the neutron distribution radius of cesium we find $\sin^2\!\vartheta_{W} =0.2396^{+0.0020}_{-0.0019}$, which should supersede the previous value determined from atomic parity violation on cesium. When including electroweak only, but also indirect, determinations of the neutron distribution radius of cesium the uncertainty reduces to 0.0017 maintaining the same central value, showing an excellent agreement independently of the method used.

\end{abstract}

\maketitle

\section{Introduction}
\label{sec:intro}

The standard model (SM) of the electroweak interactions is described by the gauge group SU(2)$\times$U(1), with the $i= 1,2,3$ gauge bosons $W^i_\mu$ and $B_\mu$ for the SU(2) and U(1) groups, respectively, and the corresponding gauge coupling constants $g$ and $g^\prime$. After spontaneous symmetry breaking, the physical $Z$ boson and photon mediators are obtained from a rotation of the basis of the two gauge bosons, $B_\mu$ and $W_\mu^3$. The angle of this rotation is known as the weak mixing angle, $\vartheta_{\text{W}}\equiv \mathrm{tan}^{-1}(g^\prime/g)$, also referred to as the Weinberg angle~\cite{ParticleDataGroup:2022pth}. In practice, the quantity \wma is usually quoted instead of the weak mixing angle itself.

The experimental determination of \wma and its dependence on the energy scale of the process, so-called running, provides a direct probe of physics phenomena beyond the SM (BSM). Its value is
extracted from neutral-current processes and $Z$-pole observables. More in detail, at the LEP Collider~\cite{ALEPH:2005ab}, it was possible to achieve the most precise measurements of \wma in the high-energy electroweak (EW) sector, in perfect agreement with other collider determinations~\cite{ParticleDataGroup:2022pth} (Tevatron, LHC and SLC). In the mid-energy range, the most precise result has been derived from the measurement of the weak charge of the proton, $Q^{p}_{W}$, performed by the $Q_{\text{weak}}$ Collaboration and found to be $Q^{p}_{W}=0.0719\pm0.0045$~\cite{Qweak:2018tjf}, showing an excellent agreement with the predicted SM running.
Moving to the low-energy sector~\cite{Kumar:2013yoa}, the most precise weak mixing angle measurement so far belongs to the so-called atomic parity violation (APV) experiments, also known as parity nonconservation (PNC), using caesium atoms~\cite{Wood:1997zq,Dzuba:2012kx}, namely $\sin^2\!\vartheta_{W} =0.2367\pm0.0018$. This value is slightly smaller than the SM prediction at near zero momentum transfer, $\mathrm{Q}=0$, calculated in the so-called modified minimal subtraction ($\overline{ MS }$) renormalization scheme, \mbox{$\sin^2\!\vartheta_W^{\rm SM} (\mathrm{Q}=0) = 0.23863\pm0.00005$}~\cite{ParticleDataGroup:2022pth, Erler:2004in,Erler:2017knj}. Atomic parity violation is caused by the weak interaction, and it is manifested in P-violating atomic observables~\cite{Roberts_2015}. Other targets have also been used, even if with less precise outcomes, and of interest for this work is the measurement of APV in lead~\cite{PhysRevA.93.012501,PhysRevLett.71.3442}. Such experiments play a unique role complementary to those at high-energy~\cite{Safronova:2017xyt}. In particular, APV is highly sensitive to extra light $Z'$ bosons predicted by BSM theories, underscoring the need for improved experimental determinations of \wma in the low-energy regime~\cite{Safronova:2017xyt,Cadeddu:2021dqx}.

\begin{figure}[!t]
\centering
\includegraphics[width=\linewidth]{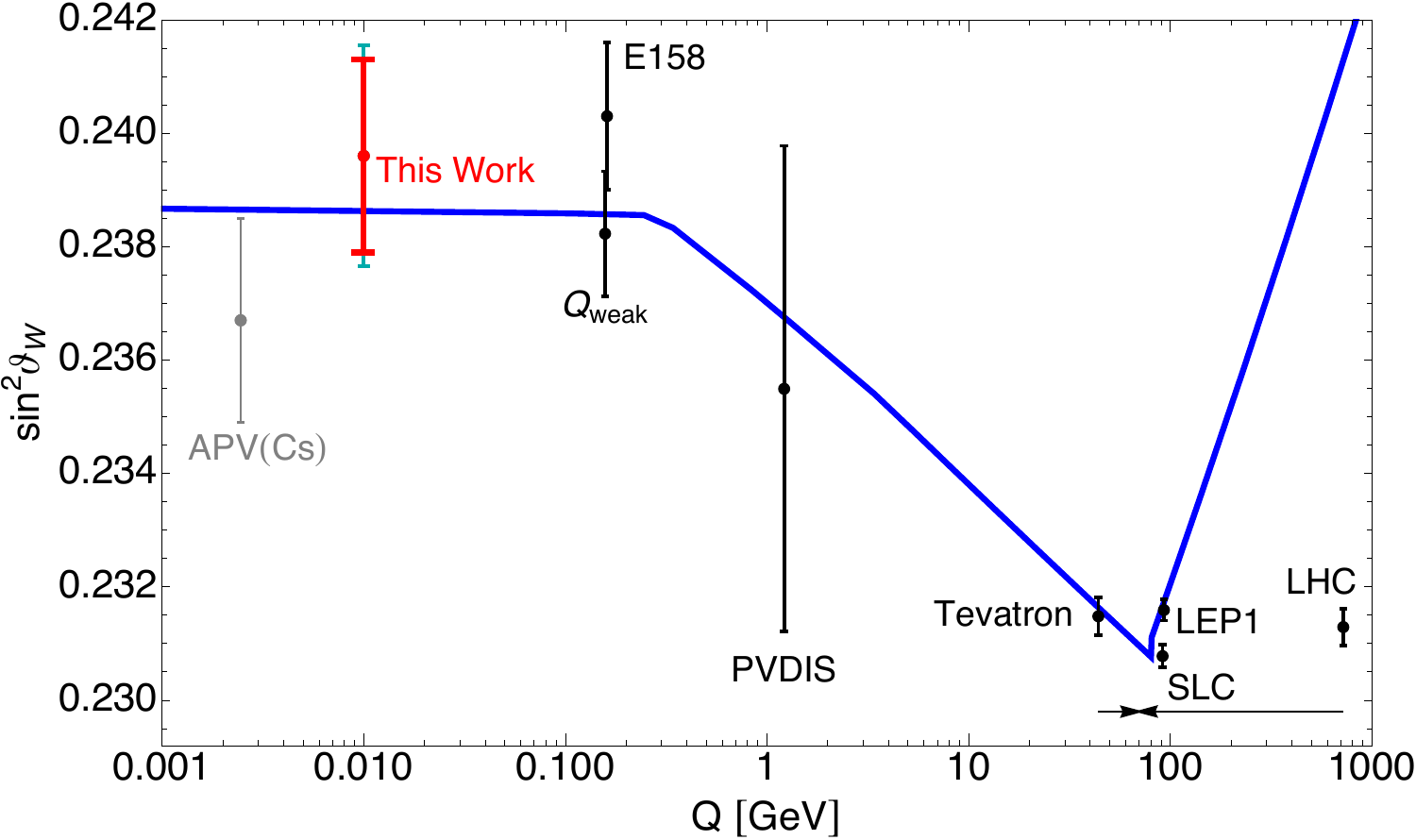}
\caption{ \label{fig:running}
Variation of $\sin^2\!\vartheta_{\text{W}}$ with scale Q. The SM prediction is shown as the solid curve, together with
experimental determinations in black at the $Z$-pole~\cite{ParticleDataGroup:2022pth} (Tevatron, LEP1, SLC, LHC),
from APV on cesium~\cite{Wood:1997zq,Dzuba:2012kx} (APV(Cs)), M{\o}ller scattering~\cite{SLACE158:2005uay} (E158), deep inelastic scattering of polarized electrons on deuterons~\cite{PVDIS:2014cmd} (PVDIS)  and the result from the proton's weak charge~\cite{Qweak:2018tjf} ($ Q_\text{weak} $). For illustration purposes, the
Tevatron and LHC points have been shifted horizontally to the left and right, respectively.  In cyan the result derived in this paper when combining APV(Cs) with \cenns COHERENT CsI data~\cite{COHERENT:2017ipa,COHERENT:2021xmm} and the neutron skin determination at the CSRe facility~\cite{Huang:2024jbh}. A similar result, slightly more precise, is obtained considering electroweak probes only and is shown in red.}
\end{figure}

A summary of the most precise weak mixing angle measurements as a function of the scale, Q, is shown in Fig.~\ref{fig:running}, along with the SM predicted running of $\sin^2\!\vartheta_{\text{W}}$, calculated in the $\overline{ MS }$ renormalization scheme~\cite{ParticleDataGroup:2022pth, Erler:2004in,Erler:2017knj}.

In the low-energy sector, there are two other electroweak probes mildly sensitive to the weak mixing angle. They are the coherent elastic neutrino-nucleus scattering (CE$\nu$NS)~\cite{Cadeddu:2023tkp} and measurements of parity-violation in electron scattering (PVES) on nuclei.
The first process has been observed so far in three targets, namely in cesium iodide (CsI)~\cite{COHERENT:2017ipa,COHERENT:2021xmm}, in argon (Ar)~\cite{COHERENT:2020ybo} and very recently in germanium (Ge)~\cite{Adamski:2024yqt} by the COHERENT Collaboration using a spallation neutron source. Moreover, a strong preference for \cenns was reported in Ref.~\cite{Colaresi:2022obx} using antineutrinos from the Dresden-II reactor and a germanium target, even though this observation is in mild tension with recent CONUS data~\cite{Ackermann:2024kxo}, and relies on an enhancement at low-energy of the ionization yield~\cite{Collar:2021fcl}, whose origin remains still unexplained~\cite{AtzoriCorona:2023ais}
.
The CE$\nu$NS cross section depends on the value of $\sin^2\!\vartheta_{W}$
through the neutral-current vector coupling of the proton $g_{V}^{p}$~\cite{AtzoriCorona:2024rtv, AtzoriCorona:2023ktl,Cadeddu:2023tkp}, whose tree-level value is given by $g_{V}^{p} = \frac{1}{2}-2\;\sin^2\!\vartheta_W^{\rm SM} (\mathrm{Q}=0) \simeq 0.0227$. More precise values are determined by taking into account the radiative corrections in the Minimal Subtraction ($\overline{\mathrm{MS}}$) scheme, following Refs.~\cite{ParticleDataGroup:2022pth,AtzoriCorona:2024rtv}. Being the proton contribution subdominant
with respect to the neutron one, only broad constraints on \wma can be obtained~\cite{AtzoriCorona:2022qrf,Cadeddu:2021ijh,Cadeddu:2019eta,Miranda:2020tif,Papoulias:2019lfi,DeRomeri:2022twg,AtzoriCorona:2023ktl}. The most recent updated result is
$\sin^2\!\vartheta_{W} = 0.231^{+0.027}_{-0.024}$~\cite{AtzoriCorona:2023ktl},
obtained from the latest COHERENT CsI data~\cite{COHERENT:2021xmm}. \\
PVES consists of polarized electron-nucleus scattering, e.g. in lead, that happens through both the weak and the electromagnetic currents. Isolating the first contribution, it provides an interesting way to assess the nuclear structure, but it can also be used to put constraints on $\sin^2\!\vartheta_{W}$. This has been recently suggested in Ref.~\cite{Corona:2021yfd}, using the latest PVES measurements on lead released by the PREX-II Collaboration~\cite{PREX:2021umo}.\\

Historically, the APV measurement in cesium has moved significantly over the years (see the inset of Fig. 9 of Ref.~\cite{AtzoriCorona:2023ktl}), being mostly lower than the SM prediction, motivating a further investigation of all the inputs entering this measurement.
Moreover, the extraction of the weak mixing angle value using electroweak probes (APV, CE$\nu$NS and PVES) is always affected by the limited knowledge of the so-called neutron skin of the nuclei used as a target~\cite{Thiel:2019tkm}. The latter is defined as $\Delta R_{\rm{np}}\equiv R_{\rm{n}}-R_{\rm{p}}$, and quantifies the difference between the neutron and the proton root-mean-square nuclear distribution radii, $R_{\rm{n}}$ and $R_{\rm{p}}$, respectively, where the latter is experimentally well known from electromagnetic measurements~\cite{Angeli:2013epw}. The usage of an extrapolated or imprecise value of the neutron radius of cesium or lead would bias the extraction of \wma and vice-versa,  misinterpreting potential signs of BSM physics. It is thus of pivotal importance to exploit all available inputs on $\Delta R_{\rm{np}}$ and \wma in a combined measurement, in order to take advantage of possible correlations and minimize external assumptions.

The difficulty in measuring $\Delta R_{\rm{np}}$ is that the nuclear neutron distribution can be probed only by exploiting the strong or
weak forces. The effects of the weak neutral-current interactions, embodied by the weak charge of the nucleus, are known with good approximation thus making these measurements systematically clean. However, the statistical uncertainty is still quite limited. On the contrary, the results of experiments with hadron probes are more precise but their interpretation is difficult since the effects of strong-force interactions cannot be calculated with sufficient approximation and the interpretation can be done only by assuming a strong-interaction model with all its limitations~\cite{Thiel:2019tkm}. On top of that, the cesium neutron radius determination with hadronic probes has been historically experimentally challenging due to the low melting point and spontaneous ignition in air, resulting up to now for the APV \wma determination in the utilization of an extrapolated $R_{\rm{n}}(\mathrm{Cs})$ value from antiprotonic atom x-ray data~\cite{Trzcinska:2001sy}. However, recently, a new direct measurement of the cesium neutron skin, $0.12\pm0.21\,\mathrm{fm}$,
appeared~\cite{Huang:2024jbh}, obtained using proton-cesium elastic scattering at low momentum transfer and an in-ring reaction
technique at the Cooler Storage Ring (CSRe) at the Heavy Ion Research Facility in Lanzhou, which can be included in the derivation of $\sin^2\!\vartheta_{W}$. The authors employed this value to re-extract the COHERENT \wma value by fitting the \cenns CsI dataset, finding $\sin^2\!\vartheta_{W} = 0.227\pm0.028$. \\

Taking into account all these recent developments, in this work we combine all the available measurements of $R_{\rm{n}}(\mathrm{Cs})$, $R_{\rm{n}}(\mathrm{Pb})$ and \wma in a global fit to extract the most up to date and precise determination of the weak mixing angle at low energy. Moreover, to better check the consistency among the different inputs and techniques, we also compare the electroweak-only determination with the averages obtained using strong probes.

\section{Results}
\label{sec:results}

To start with, we combine all available measurements using cesium atoms, namely atomic parity violation on cesium, APV(Cs), CE$\nu$NS on CsI, referred to as COH, and the recent determination of $R_{\rm{n}}(\mathrm{Cs})$ at the CSRe facility. The APV observable is the weak charge of the nucleus $Q_{W}(\mathrm{Cs})$, which is extracted by means of the experimental determination of the ratio of the parity-violating amplitude, $E_{\mathrm{PNC}}$, the Stark vector transition polarizability, $\beta$, and by calculating theoretically $E_{\mathrm{PNC}}$ in terms of $Q_{W}$. For the latter, the Particle Data Group (PDG) uses the theoretical prediction of the PNC amplitude $({\rm Im}\, E_{\rm PNC})_{\rm th.}^{\rm w.n.s.}=(0.8995\pm0.0040)\times10^{-11}|e|a_B \frac{Q_W}{N}$ of Ref.~\cite{Dzuba:2012kx}, referred hereafter as APV PDG, where Im stands for the imaginary part, $a_B$ is the Bohr radius, $N$ is the number of neutrons in the nucleus and $|e|$ is the absolute value of the electric charge. The apex w.n.s. means that the neutron skin correction has not been already implemented, given that we want to extract this correction from the combined fit using external inputs. In this work, for the PNC amplitude, we use the more precise value recently calculated in Ref.~\cite{Sahoo:2021thl}, referred to as APV 21, which exploits a variant of the perturbed relativistic coupled-cluster theory which treats the contributions of the core, valence and excited states to the spin-independent parity violating electric dipole transition amplitude on the same footing, unlike the previous high precision calculations.
This latter result is in slight tension with that used by the PDG and equal to $({\rm Im}\, E_{\rm PNC})_{\rm th.}^{\rm w.n.s.}=(0.8931\pm0.0027)\times10^{-11}|e|a_B \frac{Q_W}{N}$. Besides being more precise, it is also in better agreement with those reported in Refs.~\cite{Tran_Tan_2022,Porsev:2010de}. For completeness, in Appendix~\ref{app:pdg} we repeat all the results reported in this work using APV PDG.

\begin{figure}[h!]
    \centering
    \includegraphics[width=0.9\columnwidth]{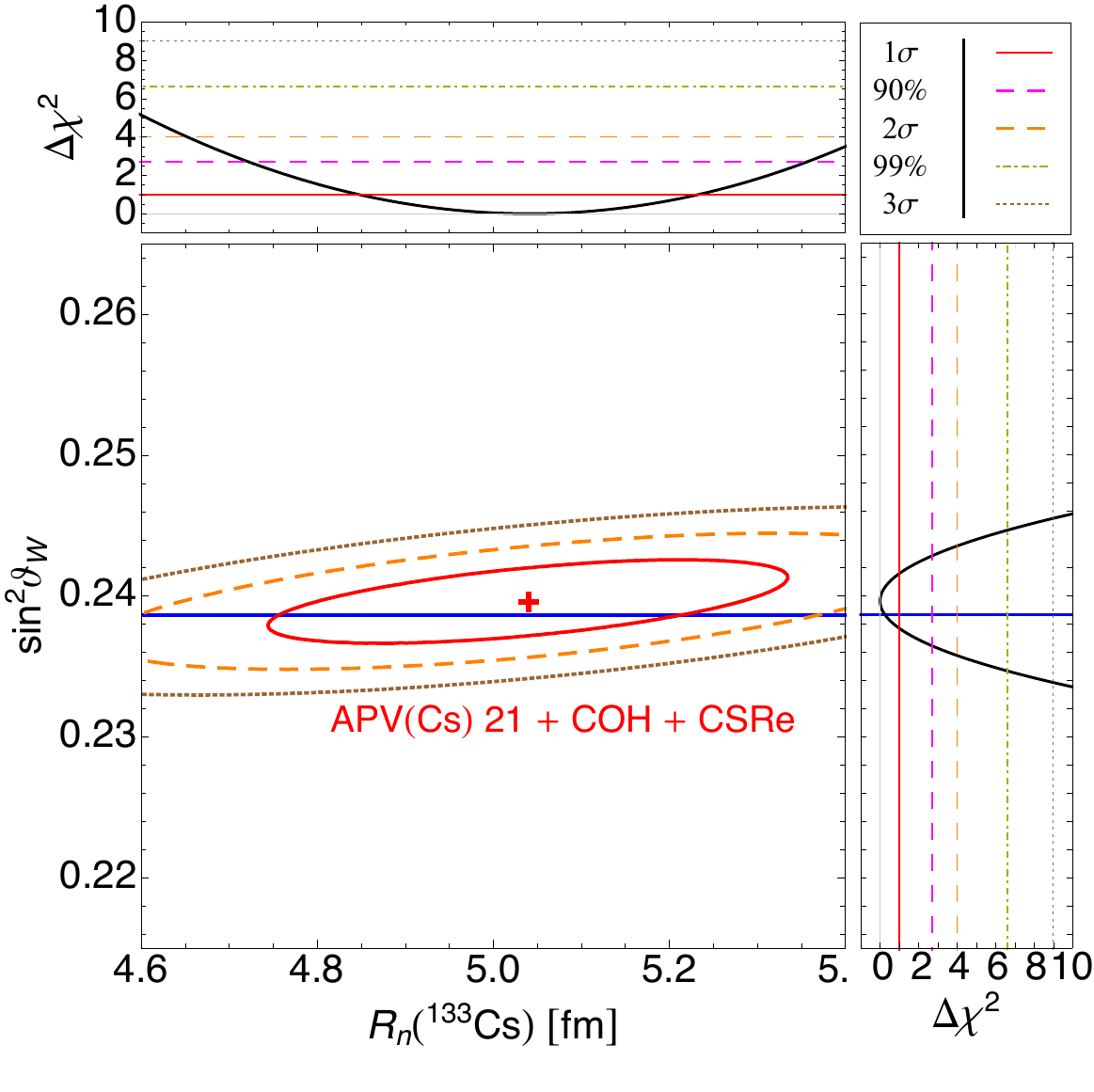}
    \caption{
    Constraints on the weak mixing angle \wma and the Cs neutron radius $R_{\rm{n}}(^{133}\rm{Cs})$ obtained from a combined APV(Cs) 21 + COH + CSRe fit at different CLs ($1-2-3\sigma$), together with their marginalizations in the side panels. The blue line indicates the theoretical low-energy value of the weak mixing angle, $\sin^2\!\vartheta_W^{\rm SM} (\mathrm{Q}=0)$.
    \label{fig:combCSRe}}
\end{figure}

\begin{table}
{\renewcommand{\arraystretch}{1.2}
\begin{tabular}{l|c|c}
& $\sin^2\vartheta_W$ & $R_{\rm{n}}(^{133}\rm{Cs}) [fm]$\\
\hline
APV(Cs)+COH+CSRe & $0.2396^{+0.0020}_{-0.0019} $  & $5.04\pm0.19$ \\[0.5mm]
EW combined & $0.2396\pm0.0017$ & $5.04\pm0.06$ \\[0.5mm]
Global fit & $0.2387\pm0.0016$  & $4.952\pm0.009$ \\[0.5mm]
\end{tabular}
}
\caption{Summary of the constraints at 1$\sigma$ CL obtained in this work on the weak mixing angle \wma and on the Cs neutron radius $R_{\rm{n}}(^{133}\rm{Cs})$. The different labels refer to the COHERENT CsI data (COH), APV (Cs) data using the PNC amplitude of
Ref.~\cite{Sahoo:2021thl}, and the CSRe determination of $R_{\rm{n}}(^{133}\rm{Cs})$. The electroweak result (EW combined) combines APV(Cs)+COH with PREX-II and APV determinations on lead. The global fit includes all of the above plus the non-EW determinations of $R_{\rm{n}}$ on lead.}  \label{tab:tablimits}
\end{table}

To combine APV(Cs) and COHERENT CsI we follow the technique initially developed in Ref.~\cite{Cadeddu:2018izq} and the latest prescriptions detailed in Ref.~\cite{AtzoriCorona:2023ktl}, adding a prior on $R_{\rm{n}}(\mathrm{Cs})=4.94 \pm 0.21\,\mathrm{fm}$\footnote{The latter radius has been obtained starting from the skin measured in Ref.~\cite{Huang:2024jbh} and using the rms proton radius of cesium $R_{\rm{p}}(\rm{Cs})=4.821(5)\, \mathrm{fm}$~\cite{Fricke:1995zz,Angeli:2013epw,AtzoriCorona:2023ktl} and the neutron radius $\langle r_{\rm{n}}^2\rangle\simeq \langle r_{\rm{p}}^2\rangle =0.708\, \mathrm{fm}^2$~\cite{Cadeddu:2020lky}.} coming from CSRe. The only mild assumptions behind this combination are that \wma is constant between the corresponding experimental momentum transfers, $2.4\lesssim Q\lesssim 100\ \mathrm{MeV}$, which is true in the absence of BSM effects, and that the $^{133}\mathrm{Cs}$ and $^{127}\mathrm{I}$ neutron skins are the same in order to isolate the contribution of $R_{\rm{n}}(\mathrm{Cs})$ when analysing the COHERENT data. Given the fact that the neutron skin difference for these two nuclei is expected to be small compared to the current precision of experimental data, this choice is a fair approximation.
We also checked that fitting for an average value of the rms neutron radii of $^{133}\mathrm{Cs}$ and $^{127}\mathrm{I}$ gives the same output. The result is shown in Fig.~\ref{fig:combCSRe} at different confidence levels (CLs), while the numerical values can be found in Tab.~\ref{tab:tablimits}.\\

\begin{figure}[t]
    \centering
    \includegraphics[width=0.9\columnwidth]{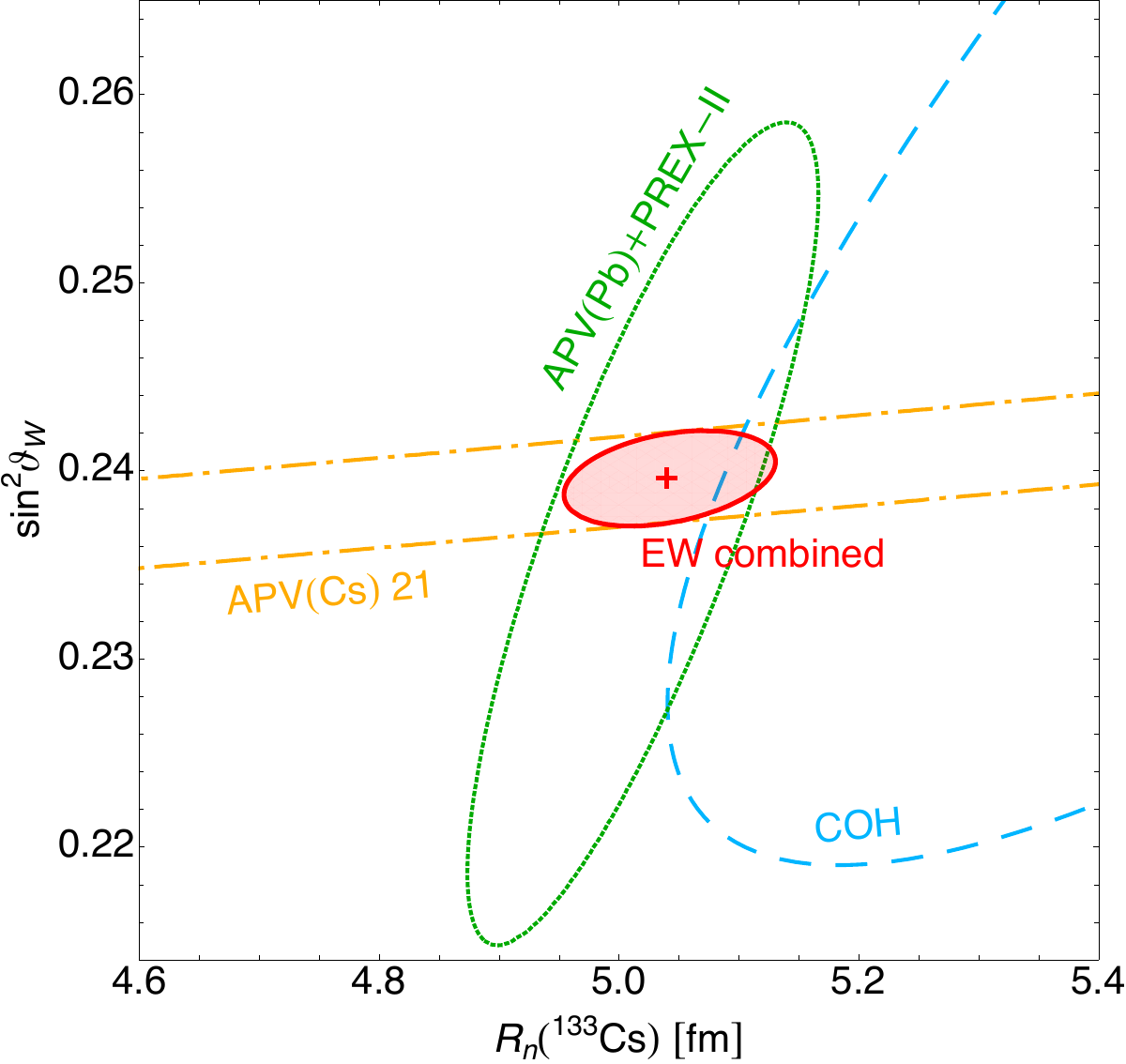}
    \caption{Individual and combined contours at $1\sigma$ CL of the available electroweak probes. Namely, APV(Cs) (orange dash-dotted line), APV(Pb)+PREX-II already converted into $R_{\rm{n}}(\mathrm{Cs})$ (dotted green line), and COH CsI (light-blue dashed line). The red solid contour is the combination of all these EW probes, with the red cross indicating the best-fit values. \label{fig:ew}}
\end{figure}

This determination of \wma depends on the CSRe determination of $R_{\rm{n}}(\mathrm{Cs})$, which dominates the COHERENT
and APV(Cs) sensitivity on the neutron distribution radius. In order to check the impact of relying so heavily on a measurement of the neutron distribution radius obtained using strong probes, we perform a further determination of \wma using electroweak probes only. To do so, we exploit two additional EW probes, namely PREX-II and APV on lead. The former determines the weak form factor value at the experimental mean momentum transfer, $Q_\mathrm{PREX-II}\simeq 78\ \mathrm{MeV}$~\cite{PREX:2021umo}, which depends on both the neutron distribution radius of lead and $\sin^2{\vartheta_{\text{W}}}$. A simultaneous fit of these two parameters can be achieved following the method developed in Ref.~\cite{Corona:2021yfd} and produces an almost fully degenerate oblique
band in the $R_{\rm{n}}(\mathrm{Pb})-\sin^2{\vartheta_{\text{W}}}$ plane. To break the degeneracy, the PREX-II result can be combined with the APV experiment on $^{208}\mathrm{Pb}$, which is sensitive to the nuclear weak charge at a momentum transfer of $Q_\mathrm{APV(\rm{Pb})}\sim 8\ \mathrm{MeV}$. In this case, we assume \wma to be constant between the corresponding experimental momentum transfers, $8\lesssim Q\lesssim 78\ \mathrm{MeV}$. Furthermore, a practically model-independent extrapolation can be performed, following the method developed in Ref.~\cite{Cadeddu:2021dqx} and briefly summarised in Appendix~\ref{app:corr}, to translate the $\mathrm{R}_{\rm{n}}(\mathrm{Pb})$ determination into a measurement of $\mathrm{R}_{\rm{n}}(\mathrm{Cs})$. In this way, the green contour shown in Fig.~\ref{fig:ew} is obtained at the 1$\sigma$ CL. In the same figure, it is possible to judge the good agreement between the different EW probes available nowadays, namely APV(Cs), APV(Pb)+PREX-II, and COHERENT CsI. All these probes can be combined together to get a fully EW determination of \wma and $\mathrm{R}_{\rm{n}}(\mathrm{Cs})$, as shown by the red contour at 1$\sigma$ in Fig.~\ref{fig:ew}.

\begin{figure}[h!]
    \centering
    \includegraphics[width=0.9\columnwidth]{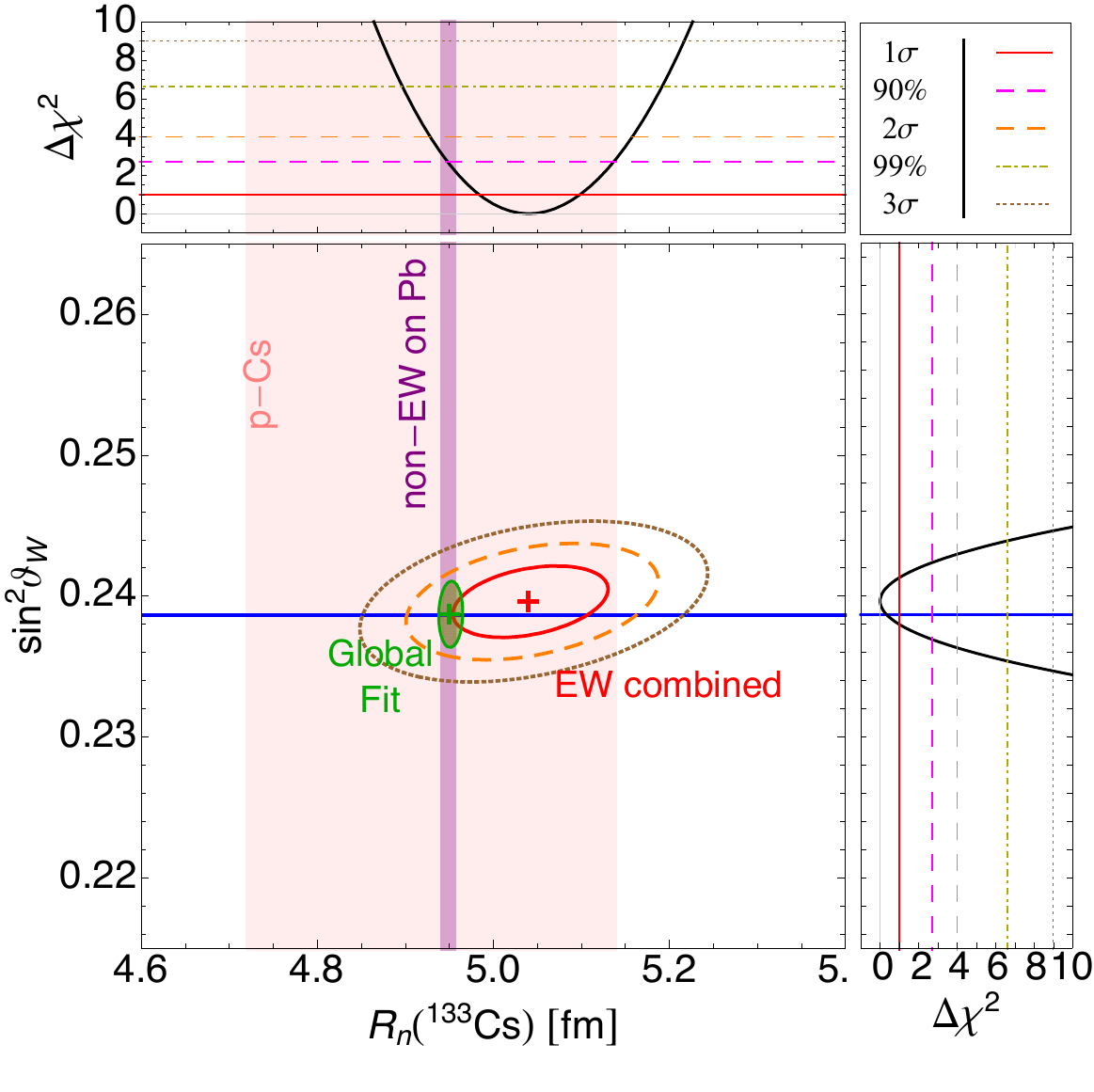}
    \caption{Combined EW fit at different CLs ($1-2-3\sigma$) and its marginalized $\Delta \chi^2$'s curves in the side panels considering APV(Cs) 21. The pink and purple bands indicate non-EW measurements of the cesium radius, coming from proton scattering on cesium (CSRe), and from the average of non-EW measurements on lead converted into $\mathrm{R}_{\rm{n}}(\mathrm{Cs})$ using the method explained in the Appendix~\ref{app:corr}. In green the result of the combined APV(Cs)21+COH+CSRe+PREX-II+APV(Pb)+non-EW(on Pb) fit is shown at $1\sigma$ CL. The blue line shows the SM value of the weak mixing angle, $\sin^2\!\vartheta_W^{\rm SM} (\mathrm{Q}=0)$.\label{fig:EWcomparison}}
\end{figure}
\begin{figure}[t]
    \centering
    \includegraphics[width=0.9\columnwidth]{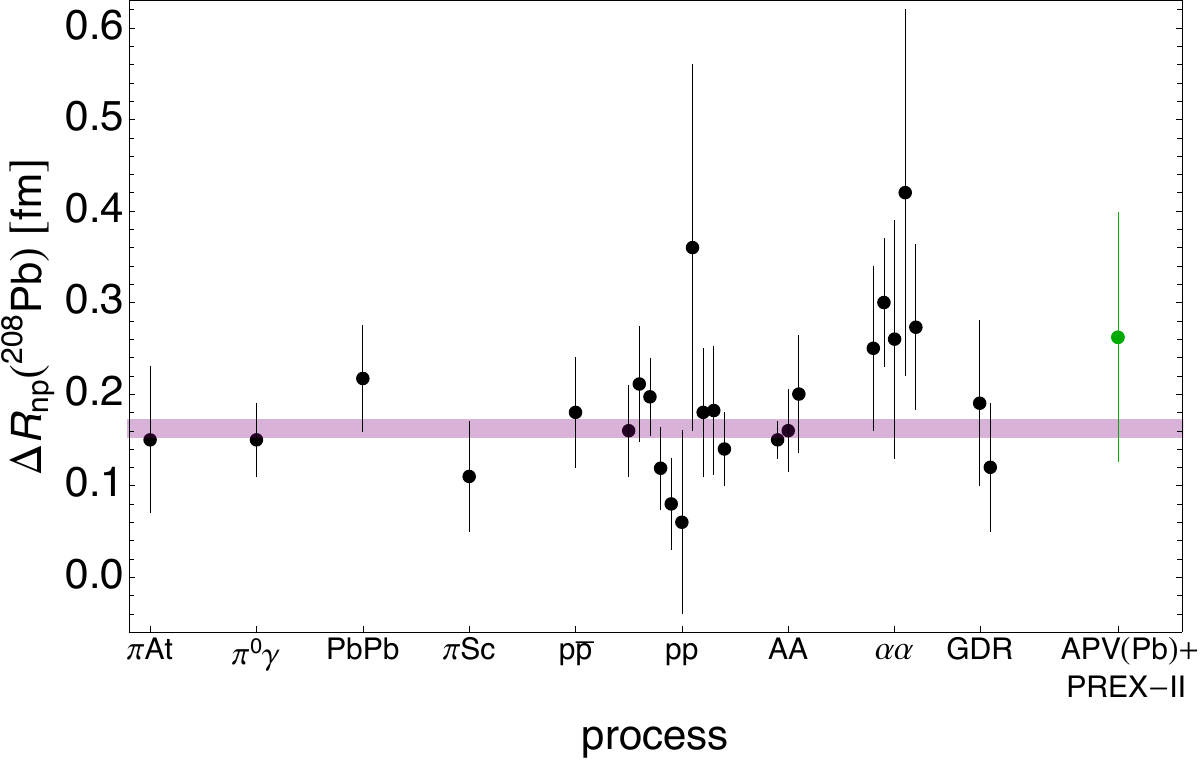}
    \caption{Summary of available measurements of the lead neutron skin considering different processes~\cite{Giacalone:2023cet,Zhang:2021jwh,Lattimer:2023xjm,Corona:2021yfd}. The black data points indicate non-EW probes, while in green we report the combined APV(Pb)+PREX-II EW measurement. The purple band indicates the $1\sigma$ CL value obtained by averaging over all the available non-EW measurements.\label{fig:rnpbnonew}}
\end{figure}

The EW combination is also shown at different CLs and together with the marginalized $\Delta \chi^2$'s curves in Fig.~\ref{fig:EWcomparison}.
Here, we also compare the EW fit to other non-EW measurements of $R_{\rm{n}}(\mathrm{Cs})$, namely the direct one derived at CSRe using proton-cesium scattering~\cite{Huang:2024jbh} and a more precise determination that is obtained from a conversion of the non-EW measurements of $R_{\rm{n}}(\mathrm{Pb})$. The latter average is retrived by considering all non-electroweak $R_{\rm{n}}(\mathrm{Pb})$ determinations in Tab. 4 of Ref.~\cite{Lattimer:2023rpe}, Tab. I of Ref.~\cite{PhysRevC.104.034303} and a recent measurement performed at the LHC~\cite{PhysRevLett.131.202302}. A summary of all the measurements considered is shown in Fig.~\ref{fig:rnpbnonew}, where it is possible to see that a rather good agreement among all the different techniques is obtained.
The average is $\Delta R^{\mathrm{non-EW}}_{{\rm{np}}}(\mathrm{Pb})=0.16\pm0.01~\mathrm{fm}$.
For comparison, in the same figure, we also show the EW determination coming from our combined fit of APV(Pb)+PREX-II, namely $\Delta R^{\mathrm{EW}}_{{\rm{np}}}(\mathrm{Pb})=0.262\pm0.136~\mathrm{fm}$, which also underlines the rather good agreement that can be obtained between EW and non-EW probes as long as in the former the dependence on the \wma is taken into account. In addition, all these experimental values are in rather good agreement with the theoretical expected range $0.13<\Delta R_{{\rm{np}}}(\mathrm{Pb})<0.19~\mathrm{[fm]}$~\cite{PhysRevC.85.041302,PhysRevC.92.064304,PhysRevC.104.024329} and the first ab-initio estimate $0.14<\Delta R_{{\rm{np}}}(\mathrm{Pb})<0.20~\mathrm{[fm]}$~\cite{Hu:2021trw}.

Interestingly, there is an extraordinary agreement among the central values obtained for \wma when using the strong probe determination at CSRe of $R_{{\rm{n}}}(\mathrm{Cs})$ (APV(Cs)+COH+CSRe) and that obtained using exclusively EW probes, as visible in Tab.~\ref{tab:tablimits}, with the latter being slightly more precise. This achievement gives confidence that an overall excellent agreement is emerging among EW and strong probe determinations of $R_{\rm{n}}$, such that the \wma result obtained is not too sensitive to the particular dataset or method used. These two determinations, which overlap besides the uncertainty, have also been added to  Fig.~\ref{fig:running} and should be compared with the previous PDG determination of APV(Cs)~\cite{ParticleDataGroup:2022pth}, the shift being largely due to the different PNC amplitude used. We would like to underline that our two determinations have uncertainties that are comparable with that of the PDG one.

For completeness, we checked what happens if a global fit of all the measurements shown in Fig.~\ref{fig:EWcomparison} is performed and it is indicated by the green contour in the same figure. Numerical values are also listed in Tab.~\ref{tab:tablimits}.
Clearly, the central value of \wma is dominated by the non-EW determinations of $R_{{\rm{n}}}(\mathrm{Pb})$, but it is possible to see that there is no much gain on the uncertainty, being the latter dominated by the uncertainty of APV(Cs). Thus, there is no clear advantage of performing such an aggressive global fit.

\section{Conclusions}
\label{sec:conclusions}

Motivated by the lack of a precise determination of the weak mixing angle at low energies we thoroughly investigate how to exploit correlations among the different probes available in order to maximize the reliability and significance of the \wma value that is extracted. In particular, we combine atomic parity violation experiments on cesium and lead nuclei, coherent elastic neutrino-nucleus scattering on cesium iodide and parity-violating electron scattering on lead by performing a fit that keeps also into account the unavoidable dependence on the experimentally poorly known neutron distribution radius of the nuclei employed. For the latter, we also exploit
a recent measurement of the cesium neutron distribution radius, obtained using proton-cesium elastic scattering at the CSRe facility. To check the consistency of the results obtained, we compare the weak-mixing angle values obtained using electroweak-only determinations of the neutron distribution radius on cesium and lead nuclei (EW combined) with that obtained using cesium-only determinations of $R_{\rm{n}}(\mathrm{Cs})$ including also strong probes (APV(Cs)+COH+CSRe). Respectively, we find
\[
\sin^2\!\vartheta_{W} =
\left\{
    \begin{array}{l}
        0.2396\pm0.0017\,(\mathrm{EW\, combined}) \\
        \, \\
        0.2396^{+0.0020}_{-0.0019}\,(\mathrm{APV(Cs)+COH+CSRe})
    \end{array}
\right.
\]
where an excellent agreement is visible, with the first method giving a slightly more precise result. These findings underscore the fact that an overall consistent picture is emerging between the values extracted using EW and strong probes, as long as the correlation with the neutron skin is properly taken into account. Finally, given that the latter combination of APV(Cs), COHERENT CsI and $R_{\rm{n}}(\mathrm{Cs})$ from CSRe uses direct determinations of $R_{\rm{n}}(\mathrm{Cs})$ it should supersede
the \wma value obtained exploiting APV(Cs)-only with an indirect extrapolation of $R_{\rm{n}}(\mathrm{Cs})$~\cite{ParticleDataGroup:2022pth}.

\begin{acknowledgements}
The work of C. Giunti is supported by the PRIN 2022 research grant ``Addressing systematic uncertainties in searches for dark matter", Number 2022F2843L, funded by MIUR.
\end{acknowledgements}



\appendix
\section{Nuclear model predictions for $\mathbf{\Delta \mathrm{R}_{{\rm{np}}} (\mathrm{Pb/Cs})}$}
\label{app:corr}

\begin{figure}[h]
    \centering
    \includegraphics[width=0.9\columnwidth]{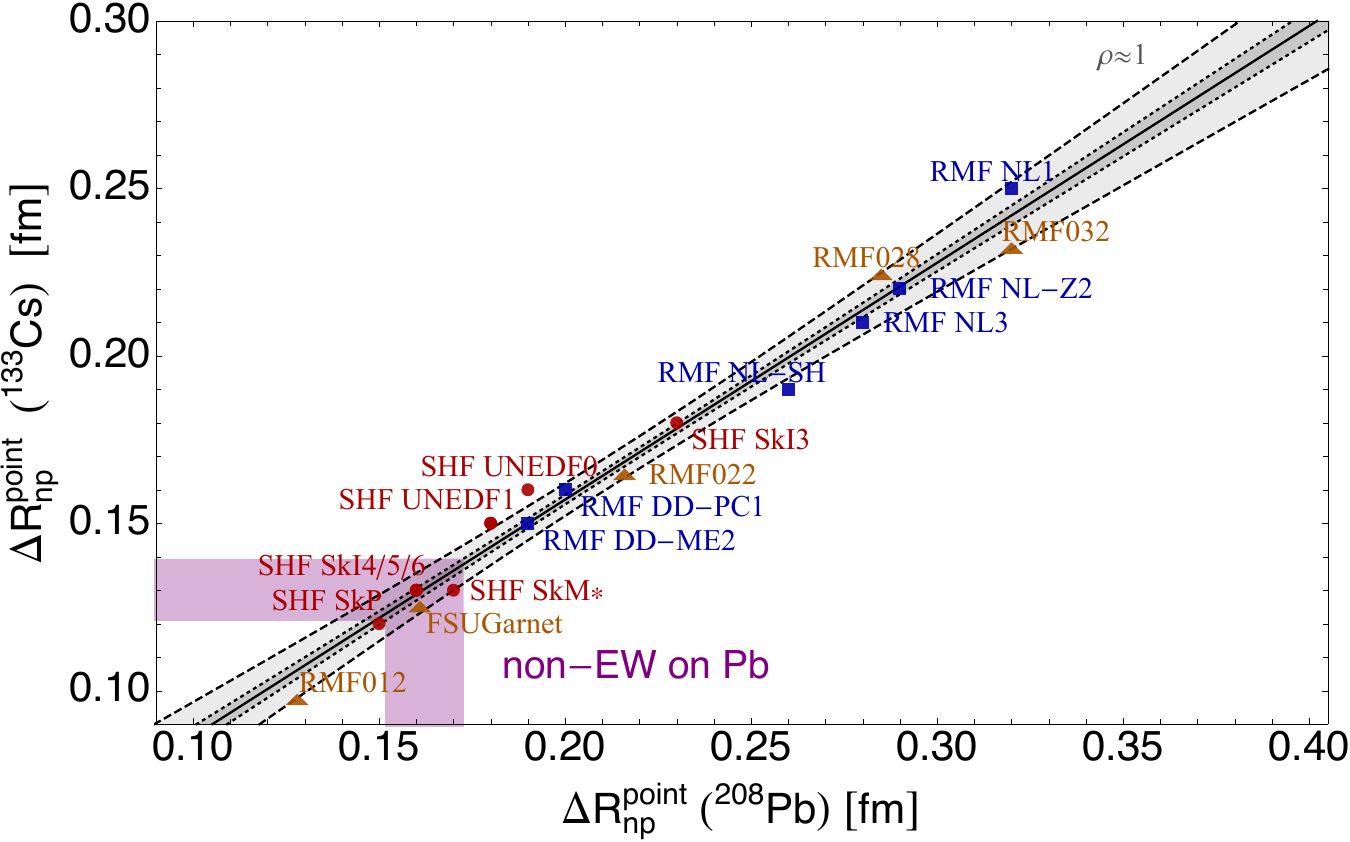}
    \caption{Correlation between the nuclear model prediction of the lead and cesium neutron skin. The purple-shaded region corresponds to the mean of the neutron skin measured through non-EW probes on lead, and its translation into cesium.\label{fig:correlation}}
\end{figure}

In Fig.~\ref{fig:correlation} we show the values of the point neutron skins\footnote{The physical proton and neutron radii
$R_{{\rm{p,n}}}$ can be retrieved
from the corresponding point-radii
$R_{{\rm{p,n}}}^{\text{point}}$
adding in quadrature the contribution of the
rms nucleon N radius
$\langle r_{{\rm{N}}}^2 \rangle^{1/2} \simeq 0.84 \, \text{fm}$,
that is considered to be approximately equal for the proton and the neutron. Namely,
$R_{{\rm{p,n}}}^2= (R_{{\rm{p,n}}}^{\text{point}})^2 + \langle r_{{\rm{N}}}^2 \rangle$.} of $^{208}\mathrm{Pb}$ and $^{133}\mathrm{Cs}$ obtained with various nonrelativistic Skyrme-Hartree-Fock (SHF)~\cite{Dobaczewski:1983zc,Bartel:1982ed,Kortelainen:2011ft,Kortelainen:2010hv,Chabanat:1997un,Reinhard:1995zz}  and relativistic mean-field (RMF)~\cite{Sharma:1993it,Bender:1999yt,Lalazissis:1996rd,Reinhard:1986qq,Niksic:2008vp,Niksic:2002yp,Hernandez:2019hsk,Yang:2019pbx,Chen:2014sca,Chen:2014mza} nuclear models. A clear model-independent linear correlation~\cite{Yang:2019pbx, Zheng:2014nga, Sil:2005tg, PhysRevC.85.041302,Yue:2021yfx,Cadeddu:2021ijh} is present between the two neutron skins within the nonrelativistic and relativistic
models with different interactions, with a Pearson’s correlation coefficient $\rho\simeq0.999$. Namely, we find $\Delta R_{{\rm{np}}}^{\mathrm{point}} (\mathrm{Cs})=0.707\times \Delta R_{{\rm{np}}} (\mathrm{Pb})+0.016$~fm that can be translated into a physical $\Delta R_{{\rm{np}}}(\mathrm{Cs})$ determination.
We
exploit this powerful linear correlation to translate the PREX-II and APV(Pb) combined measurement of the lead neutron skin into a cesium one. Moreover, we also use it to translate the mean of the neutron skin measured through non-EW probes on lead into a determination of the cesium one. The latter is shown by the purple extrapolation in Fig.~\ref{fig:correlation}.


\section{Results with Im\textit{E}$_\mathbf{\mathrm{PNC}}$ PDG}
\label{app:pdg}

\begin{figure}[h]
   \centering
    \includegraphics[width=0.8\columnwidth]{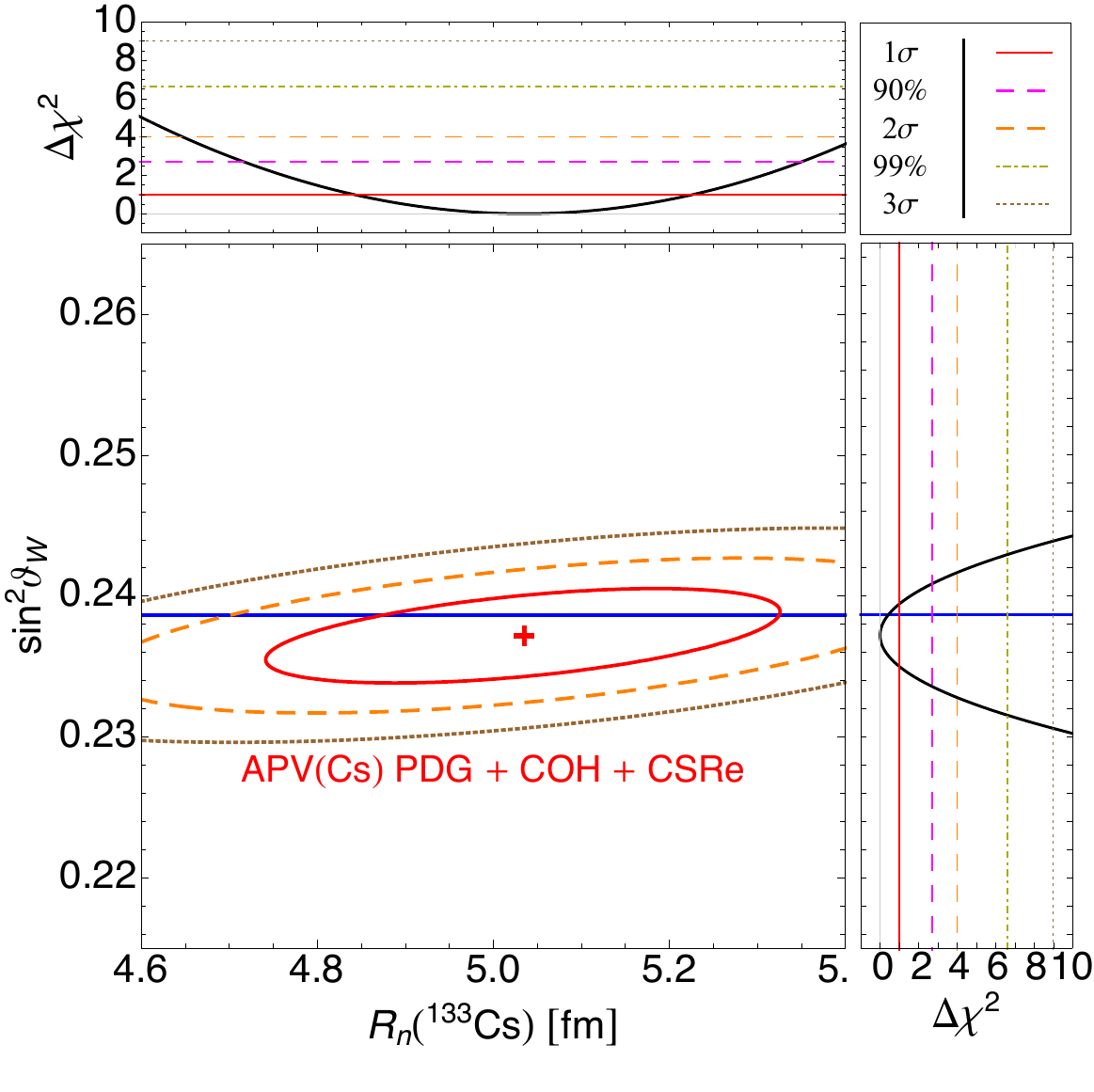}
   \caption{Constraints on the weak mixing angle \wma and the Cs neutron radius $R_{\rm{n}}(^{133}\rm{Cs})$ obtained from a combined APV (Cs) PDG + COH + CSRe fit at different CLs ($1-2-3\sigma$), together with their marginalizations in the side panels. The blue line indicates the theoretical low-energy value of the weak mixing angle, $\sin^2\!\vartheta_W^{\rm SM} (\mathrm{Q}=0)$.
   \label{fig:combCSRePDG}}
\end{figure}

In this appendix, we report the results using for the atomic parity violation determination in cesium the PNC amplitude $E_{\mathrm{PNC}}$ from Ref.~\cite{Dzuba:2012kx}, referred to as APV(Cs) PDG.
The combination of APV(Cs) PDG and COHERENT CsI adding a prior on $R_{\rm{n}}(\mathrm{Cs})=4.94 \pm 0.21\,\mathrm{fm}$ coming from CSRe is shown in Fig.~\ref{fig:combCSRePDG} at different CLs, while
the numerical values can be found in Tab.~\ref{tab:tablimitsAPVPDG}.\\

\begin{figure}[h!]
    \centering
    \includegraphics[width=0.8\columnwidth]{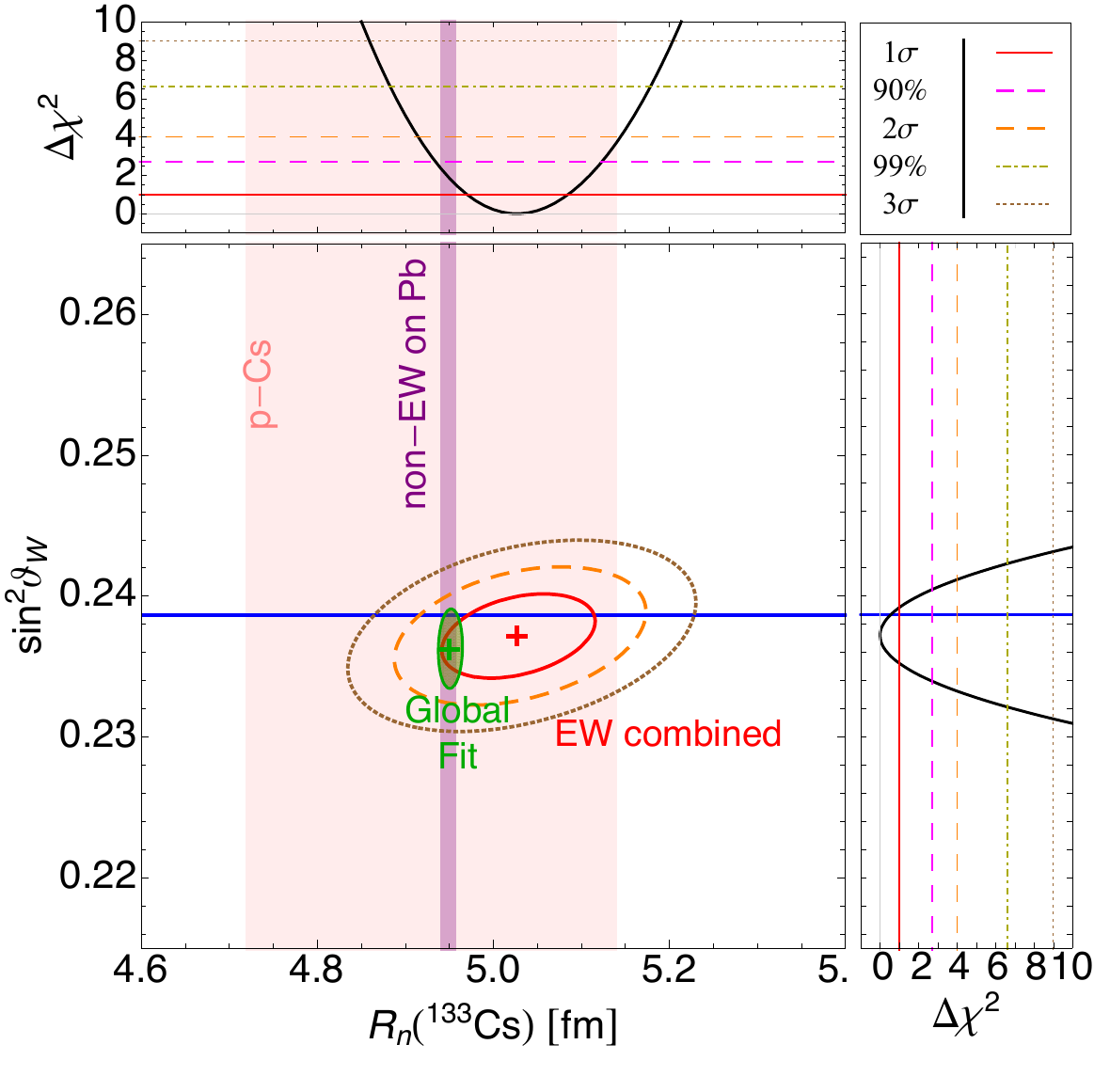}
    \caption{Combined EW fit at different CLs ($1-2-3\sigma$) and its marginalized $\Delta \chi^2$'s curves in the side panels considering APV(Cs) PDG. The pink and purple bands indicate non-EW measurements of the cesium radius, coming from proton scattering on cesium (CSRe), and from the average of non-EW measurements on lead converted into $\mathrm{R}_{\rm{n}}(\mathrm{Cs})$ using the method explained in the Appendix~\ref{app:corr}. In green the result of the combined APV(Cs) PDG+COH+CSRe+PREX-II+APV(Pb)+non-EW(on Pb) fit is shown at $1\sigma$ CL. The blue line shows the SM value of the weak mixing angle, $\sin^2\!\vartheta_W^{\rm SM} (\mathrm{Q}=0)$.\label{fig:EWcomparisonPDG}}
\end{figure}

All the EW probes available nowadays, namely APV(Cs) PDG, APV(Pb)+PREX-II, and COHERENT CsI, are combined to get a fully EW determination of \wma and $R_{\rm{n}}(\mathrm{Cs})$, as shown at different CLs and together with the marginalized $\Delta \chi^2$'s curves in Fig.~\ref{fig:EWcomparisonPDG}.
Here, we also compare the EW fit to other non-EW measurements of $R_{\rm{n}}(\mathrm{Cs})$, namely the direct one derived at CSRe using proton-cesium scattering~\cite{Huang:2024jbh} and a more precise determination that is obtained from a conversion of the non-EW measurements of $R_{\rm{n}}(\mathrm{Pb})$, as explained in Appendix~\ref{app:corr}. \\
For completeness, in Fig.~\ref{fig:EWcomparisonPDG} we also show the result of a global fit of all these determinations. It is possible to notice, by comparing Fig.~\ref{fig:EWcomparisonPDG} and Fig.~\ref{fig:EWcomparison}, that using the PNC amplitude $E_{\mathrm{PNC}}$ from Ref.~\cite{Dzuba:2012kx} (APV(Cs) PDG) results in a \wma value that is slightly smaller than the SM prediction as well as less precise.

\begin{table}[h!]
{\renewcommand{\arraystretch}{1.2}
\begin{tabular}{l|c|c}
& $\sin^2\vartheta_W$ & $R_{\rm{n}}(^{133}\rm{Cs}) [fm]$\\
\hline
APV(Cs)+COH+CSRe & $0.2372\pm0.0022 $  & $5.04\pm0.19$ \\[0.5mm]
EW combined & $0.2372\pm0.002$ & $5.03\pm0.06$ \\[0.5mm]
Global fit & $0.2363^{+0.0018}_{-0.0019}$  & $4.951\pm0.009$ \\[0.5mm]
\end{tabular}
}
\caption{Summary of the constraints at 1$\sigma$ CL obtained in this work on the weak mixing angle \wma and on the Cs neutron radius $R_{\rm{n}}(^{133}\rm{Cs})$. The different labels refer to the COHERENT CsI data (COH), APV (Cs) PDG data using the PNC amplitude of
Ref.~\cite{Dzuba:2012kx}, and the CSRe determination of $R_{\rm{n}}(^{133}\rm{Cs})$. The electroweak result (EW combined) combines APV(Cs) +COH with PREX-II and APV determinations on lead. The global fit includes all of the above plus the non-EW determinations of $R_{\rm{n}}$ on lead.
}  \label{tab:tablimitsAPVPDG}
\end{table}

\newpage

\bibliography{sample}

\begin{thebibliography}{73}%
\makeatletter
\providecommand \@ifxundefined [1]{%
 \@ifx{#1\undefined}
}%
\providecommand \@ifnum [1]{%
 \ifnum #1\expandafter \@firstoftwo
 \else \expandafter \@secondoftwo
 \fi
}%
\providecommand \@ifx [1]{%
 \ifx #1\expandafter \@firstoftwo
 \else \expandafter \@secondoftwo
 \fi
}%
\providecommand \natexlab [1]{#1}%
\providecommand \enquote  [1]{``#1''}%
\providecommand \bibnamefont  [1]{#1}%
\providecommand \bibfnamefont [1]{#1}%
\providecommand \citenamefont [1]{#1}%
\providecommand \href@noop [0]{\@secondoftwo}%
\providecommand \href [0]{\begingroup \@sanitize@url \@href}%
\providecommand \@href[1]{\@@startlink{#1}\@@href}%
\providecommand \@@href[1]{\endgroup#1\@@endlink}%
\providecommand \@sanitize@url [0]{\catcode `\\12\catcode `\$12\catcode `\&12\catcode `\#12\catcode `\^12\catcode `\_12\catcode `\%12\relax}%
\providecommand \@@startlink[1]{}%
\providecommand \@@endlink[0]{}%
\providecommand \url  [0]{\begingroup\@sanitize@url \@url }%
\providecommand \@url [1]{\endgroup\@href {#1}{\urlprefix }}%
\providecommand \urlprefix  [0]{URL }%
\providecommand \Eprint [0]{\href }%
\providecommand \doibase [0]{https://doi.org/}%
\providecommand \selectlanguage [0]{\@gobble}%
\providecommand \bibinfo  [0]{\@secondoftwo}%
\providecommand \bibfield  [0]{\@secondoftwo}%
\providecommand \translation [1]{[#1]}%
\providecommand \BibitemOpen [0]{}%
\providecommand \bibitemStop [0]{}%
\providecommand \bibitemNoStop [0]{.\EOS\space}%
\providecommand \EOS [0]{\spacefactor3000\relax}%
\providecommand \BibitemShut  [1]{\csname bibitem#1\endcsname}%
\let\auto@bib@innerbib\@empty
\bibitem [{\citenamefont {Workman}\ \emph {et~al.}(2022)\citenamefont {Workman} \emph {et~al.}}]{ParticleDataGroup:2022pth}%
  \BibitemOpen
  \bibfield  {author} {\bibinfo {author} {\bibfnamefont {R.~L.}\ \bibnamefont {Workman}} \emph {et~al.} (\bibinfo {collaboration} {Particle Data Group}),\ }\bibfield  {title} {\bibinfo {title} {{Review of Particle Physics}},\ }\href {https://doi.org/10.1093/ptep/ptac097} {\bibfield  {journal} {\bibinfo  {journal} {PTEP}\ }\textbf {\bibinfo {volume} {2022}},\ \bibinfo {pages} {083C01} (\bibinfo {year} {2022})}\BibitemShut {NoStop}%
\bibitem [{\citenamefont {Schael}\ \emph {et~al.}(2006)\citenamefont {Schael} \emph {et~al.}}]{ALEPH:2005ab}%
  \BibitemOpen
  \bibfield  {author} {\bibinfo {author} {\bibfnamefont {S.}~\bibnamefont {Schael}} \emph {et~al.} (\bibinfo {collaboration} {ALEPH, DELPHI, L3, OPAL, SLD, LEP Electroweak Working Group, SLD Electroweak Group, SLD Heavy Flavour Group}),\ }\bibfield  {title} {\bibinfo {title} {{Precision electroweak measurements on the $Z$ resonance}},\ }\href {https://doi.org/10.1016/j.physrep.2005.12.006} {\bibfield  {journal} {\bibinfo  {journal} {Phys. Rept.}\ }\textbf {\bibinfo {volume} {427}},\ \bibinfo {pages} {257} (\bibinfo {year} {2006})},\ \Eprint {https://arxiv.org/abs/hep-ex/0509008} {arXiv:hep-ex/0509008} \BibitemShut {NoStop}%
\bibitem [{\citenamefont {Androi\'c}\ \emph {et~al.}(2018)\citenamefont {Androi\'c} \emph {et~al.}}]{Qweak:2018tjf}%
  \BibitemOpen
  \bibfield  {author} {\bibinfo {author} {\bibfnamefont {D.}~\bibnamefont {Androi\'c}} \emph {et~al.} (\bibinfo {collaboration} {Qweak}),\ }\bibfield  {title} {\bibinfo {title} {{Precision measurement of the weak charge of the proton}},\ }\href {https://doi.org/10.1038/s41586-018-0096-0} {\bibfield  {journal} {\bibinfo  {journal} {Nature}\ }\textbf {\bibinfo {volume} {557}},\ \bibinfo {pages} {207} (\bibinfo {year} {2018})},\ \Eprint {https://arxiv.org/abs/1905.08283} {arXiv:1905.08283 [nucl-ex]} \BibitemShut {NoStop}%
\bibitem [{\citenamefont {Kumar}\ \emph {et~al.}(2013)\citenamefont {Kumar}, \citenamefont {Mantry}, \citenamefont {Marciano},\ and\ \citenamefont {Souder}}]{Kumar:2013yoa}%
  \BibitemOpen
  \bibfield  {author} {\bibinfo {author} {\bibfnamefont {K.~S.}\ \bibnamefont {Kumar}}, \bibinfo {author} {\bibfnamefont {S.}~\bibnamefont {Mantry}}, \bibinfo {author} {\bibfnamefont {W.~J.}\ \bibnamefont {Marciano}},\ and\ \bibinfo {author} {\bibfnamefont {P.~A.}\ \bibnamefont {Souder}},\ }\bibfield  {title} {\bibinfo {title} {{Low Energy Measurements of the Weak Mixing Angle}},\ }\href {https://doi.org/10.1146/annurev-nucl-102212-170556} {\bibfield  {journal} {\bibinfo  {journal} {Ann. Rev. Nucl. Part. Sci.}\ }\textbf {\bibinfo {volume} {63}},\ \bibinfo {pages} {237} (\bibinfo {year} {2013})},\ \Eprint {https://arxiv.org/abs/1302.6263} {arXiv:1302.6263 [hep-ex]} \BibitemShut {NoStop}%
\bibitem [{\citenamefont {Wood}\ \emph {et~al.}(1997)\citenamefont {Wood}, \citenamefont {Bennett}, \citenamefont {Cho}, \citenamefont {Masterson}, \citenamefont {Roberts}, \citenamefont {Tanner},\ and\ \citenamefont {Wieman}}]{Wood:1997zq}%
  \BibitemOpen
  \bibfield  {author} {\bibinfo {author} {\bibfnamefont {C.~S.}\ \bibnamefont {Wood}}, \bibinfo {author} {\bibfnamefont {S.~C.}\ \bibnamefont {Bennett}}, \bibinfo {author} {\bibfnamefont {D.}~\bibnamefont {Cho}}, \bibinfo {author} {\bibfnamefont {B.~P.}\ \bibnamefont {Masterson}}, \bibinfo {author} {\bibfnamefont {J.~L.}\ \bibnamefont {Roberts}}, \bibinfo {author} {\bibfnamefont {C.~E.}\ \bibnamefont {Tanner}},\ and\ \bibinfo {author} {\bibfnamefont {C.~E.}\ \bibnamefont {Wieman}},\ }\bibfield  {title} {\bibinfo {title} {{Measurement of parity nonconservation and an anapole moment in cesium}},\ }\href {https://doi.org/10.1126/science.275.5307.1759} {\bibfield  {journal} {\bibinfo  {journal} {Science}\ }\textbf {\bibinfo {volume} {275}},\ \bibinfo {pages} {1759} (\bibinfo {year} {1997})}\BibitemShut {NoStop}%
\bibitem [{\citenamefont {Dzuba}\ \emph {et~al.}(2012)\citenamefont {Dzuba}, \citenamefont {Berengut}, \citenamefont {Flambaum},\ and\ \citenamefont {Roberts}}]{Dzuba:2012kx}%
  \BibitemOpen
  \bibfield  {author} {\bibinfo {author} {\bibfnamefont {V.~A.}\ \bibnamefont {Dzuba}}, \bibinfo {author} {\bibfnamefont {J.~C.}\ \bibnamefont {Berengut}}, \bibinfo {author} {\bibfnamefont {V.~V.}\ \bibnamefont {Flambaum}},\ and\ \bibinfo {author} {\bibfnamefont {B.}~\bibnamefont {Roberts}},\ }\bibfield  {title} {\bibinfo {title} {{Revisiting parity non-conservation in cesium}},\ }\href {https://doi.org/10.1103/PhysRevLett.109.203003} {\bibfield  {journal} {\bibinfo  {journal} {Phys. Rev. Lett.}\ }\textbf {\bibinfo {volume} {109}},\ \bibinfo {pages} {203003} (\bibinfo {year} {2012})},\ \Eprint {https://arxiv.org/abs/1207.5864} {arXiv:1207.5864 [hep-ph]} \BibitemShut {NoStop}%
\bibitem [{\citenamefont {Erler}\ and\ \citenamefont {Ramsey-Musolf}(2005)}]{Erler:2004in}%
  \BibitemOpen
  \bibfield  {author} {\bibinfo {author} {\bibfnamefont {J.}~\bibnamefont {Erler}}\ and\ \bibinfo {author} {\bibfnamefont {M.~J.}\ \bibnamefont {Ramsey-Musolf}},\ }\bibfield  {title} {\bibinfo {title} {{The Weak mixing angle at low energies}},\ }\href {https://doi.org/10.1103/PhysRevD.72.073003} {\bibfield  {journal} {\bibinfo  {journal} {Phys. Rev. D}\ }\textbf {\bibinfo {volume} {72}},\ \bibinfo {pages} {073003} (\bibinfo {year} {2005})},\ \Eprint {https://arxiv.org/abs/hep-ph/0409169} {arXiv:hep-ph/0409169} \BibitemShut {NoStop}%
\bibitem [{\citenamefont {Erler}\ and\ \citenamefont {Ferro-Hern\'andez}(2018)}]{Erler:2017knj}%
  \BibitemOpen
  \bibfield  {author} {\bibinfo {author} {\bibfnamefont {J.}~\bibnamefont {Erler}}\ and\ \bibinfo {author} {\bibfnamefont {R.}~\bibnamefont {Ferro-Hern\'andez}},\ }\bibfield  {title} {\bibinfo {title} {{Weak Mixing Angle in the Thomson Limit}},\ }\href {https://doi.org/10.1007/JHEP03(2018)196} {\bibfield  {journal} {\bibinfo  {journal} {JHEP}\ }\textbf {\bibinfo {volume} {03}},\ \bibinfo {pages} {196}},\ \Eprint {https://arxiv.org/abs/1712.09146} {arXiv:1712.09146 [hep-ph]} \BibitemShut {NoStop}%
\bibitem [{\citenamefont {Roberts}\ \emph {et~al.}(2015)\citenamefont {Roberts}, \citenamefont {Dzuba},\ and\ \citenamefont {Flambaum}}]{Roberts_2015}%
  \BibitemOpen
  \bibfield  {author} {\bibinfo {author} {\bibfnamefont {B.}~\bibnamefont {Roberts}}, \bibinfo {author} {\bibfnamefont {V.}~\bibnamefont {Dzuba}},\ and\ \bibinfo {author} {\bibfnamefont {V.}~\bibnamefont {Flambaum}},\ }\bibfield  {title} {\bibinfo {title} {Parity and time-reversal violation in atomic systems},\ }\href {https://doi.org/10.1146/annurev-nucl-102014-022331} {\bibfield  {journal} {\bibinfo  {journal} {Annual Review of Nuclear and Particle Science}\ }\textbf {\bibinfo {volume} {65}},\ \bibinfo {pages} {63–86} (\bibinfo {year} {2015})}\BibitemShut {NoStop}%
\bibitem [{\citenamefont {Porsev}\ \emph {et~al.}(2016)\citenamefont {Porsev}, \citenamefont {Kozlov}, \citenamefont {Safronova},\ and\ \citenamefont {Tupitsyn}}]{PhysRevA.93.012501}%
  \BibitemOpen
  \bibfield  {author} {\bibinfo {author} {\bibfnamefont {S.~G.}\ \bibnamefont {Porsev}}, \bibinfo {author} {\bibfnamefont {M.~G.}\ \bibnamefont {Kozlov}}, \bibinfo {author} {\bibfnamefont {M.~S.}\ \bibnamefont {Safronova}},\ and\ \bibinfo {author} {\bibfnamefont {I.~I.}\ \bibnamefont {Tupitsyn}},\ }\bibfield  {title} {\bibinfo {title} {Development of the configuration-interaction + all-order method and application to the parity-nonconserving amplitude and other properties of pb},\ }\href {https://doi.org/10.1103/PhysRevA.93.012501} {\bibfield  {journal} {\bibinfo  {journal} {Phys. Rev. A}\ }\textbf {\bibinfo {volume} {93}},\ \bibinfo {pages} {012501} (\bibinfo {year} {2016})}\BibitemShut {NoStop}%
\bibitem [{\citenamefont {Meekhof}\ \emph {et~al.}(1993)\citenamefont {Meekhof}, \citenamefont {Vetter}, \citenamefont {Majumder}, \citenamefont {Lamoreaux},\ and\ \citenamefont {Fortson}}]{PhysRevLett.71.3442}%
  \BibitemOpen
  \bibfield  {author} {\bibinfo {author} {\bibfnamefont {D.~M.}\ \bibnamefont {Meekhof}}, \bibinfo {author} {\bibfnamefont {P.}~\bibnamefont {Vetter}}, \bibinfo {author} {\bibfnamefont {P.~K.}\ \bibnamefont {Majumder}}, \bibinfo {author} {\bibfnamefont {S.~K.}\ \bibnamefont {Lamoreaux}},\ and\ \bibinfo {author} {\bibfnamefont {E.~N.}\ \bibnamefont {Fortson}},\ }\bibfield  {title} {\bibinfo {title} {High-precision measurement of parity nonconserving optical rotation in atomic lead},\ }\href {https://doi.org/10.1103/PhysRevLett.71.3442} {\bibfield  {journal} {\bibinfo  {journal} {Phys. Rev. Lett.}\ }\textbf {\bibinfo {volume} {71}},\ \bibinfo {pages} {3442} (\bibinfo {year} {1993})}\BibitemShut {NoStop}%
\bibitem [{\citenamefont {Safronova}\ \emph {et~al.}(2018)\citenamefont {Safronova}, \citenamefont {Budker}, \citenamefont {DeMille}, \citenamefont {Kimball}, \citenamefont {Derevianko},\ and\ \citenamefont {Clark}}]{Safronova:2017xyt}%
  \BibitemOpen
  \bibfield  {author} {\bibinfo {author} {\bibfnamefont {M.~S.}\ \bibnamefont {Safronova}}, \bibinfo {author} {\bibfnamefont {D.}~\bibnamefont {Budker}}, \bibinfo {author} {\bibfnamefont {D.}~\bibnamefont {DeMille}}, \bibinfo {author} {\bibfnamefont {D.~F.~J.}\ \bibnamefont {Kimball}}, \bibinfo {author} {\bibfnamefont {A.}~\bibnamefont {Derevianko}},\ and\ \bibinfo {author} {\bibfnamefont {C.~W.}\ \bibnamefont {Clark}},\ }\bibfield  {title} {\bibinfo {title} {{Search for New Physics with Atoms and Molecules}},\ }\href {https://doi.org/10.1103/RevModPhys.90.025008} {\bibfield  {journal} {\bibinfo  {journal} {Rev. Mod. Phys.}\ }\textbf {\bibinfo {volume} {90}},\ \bibinfo {pages} {025008} (\bibinfo {year} {2018})},\ \Eprint {https://arxiv.org/abs/1710.01833} {arXiv:1710.01833 [physics.atom-ph]} \BibitemShut {NoStop}%
\bibitem [{\citenamefont {Cadeddu}\ \emph {et~al.}(2021{\natexlab{a}})\citenamefont {Cadeddu}, \citenamefont {Cargioli}, \citenamefont {Dordei}, \citenamefont {Giunti},\ and\ \citenamefont {Picciau}}]{Cadeddu:2021dqx}%
  \BibitemOpen
  \bibfield  {author} {\bibinfo {author} {\bibfnamefont {M.}~\bibnamefont {Cadeddu}}, \bibinfo {author} {\bibfnamefont {N.}~\bibnamefont {Cargioli}}, \bibinfo {author} {\bibfnamefont {F.}~\bibnamefont {Dordei}}, \bibinfo {author} {\bibfnamefont {C.}~\bibnamefont {Giunti}},\ and\ \bibinfo {author} {\bibfnamefont {E.}~\bibnamefont {Picciau}},\ }\bibfield  {title} {\bibinfo {title} {{Muon and electron g-2 and proton and cesium weak charges implications on dark Zd models}},\ }\href {https://doi.org/10.1103/PhysRevD.104.L011701} {\bibfield  {journal} {\bibinfo  {journal} {Phys. Rev. D}\ }\textbf {\bibinfo {volume} {104}},\ \bibinfo {pages} {011701} (\bibinfo {year} {2021}{\natexlab{a}})},\ \Eprint {https://arxiv.org/abs/2104.03280} {arXiv:2104.03280 [hep-ph]} \BibitemShut {NoStop}%
\bibitem [{\citenamefont {Anthony}\ \emph {et~al.}(2005)\citenamefont {Anthony} \emph {et~al.}}]{SLACE158:2005uay}%
  \BibitemOpen
  \bibfield  {author} {\bibinfo {author} {\bibfnamefont {P.~L.}\ \bibnamefont {Anthony}} \emph {et~al.} (\bibinfo {collaboration} {SLAC E158}),\ }\bibfield  {title} {\bibinfo {title} {{Precision measurement of the weak mixing angle in Moller scattering}},\ }\href {https://doi.org/10.1103/PhysRevLett.95.081601} {\bibfield  {journal} {\bibinfo  {journal} {Phys. Rev. Lett.}\ }\textbf {\bibinfo {volume} {95}},\ \bibinfo {pages} {081601} (\bibinfo {year} {2005})},\ \Eprint {https://arxiv.org/abs/hep-ex/0504049} {arXiv:hep-ex/0504049} \BibitemShut {NoStop}%
\bibitem [{\citenamefont {Wang}\ \emph {et~al.}(2014)\citenamefont {Wang} \emph {et~al.}}]{PVDIS:2014cmd}%
  \BibitemOpen
  \bibfield  {author} {\bibinfo {author} {\bibfnamefont {D.}~\bibnamefont {Wang}} \emph {et~al.} (\bibinfo {collaboration} {PVDIS}),\ }\bibfield  {title} {\bibinfo {title} {{Measurement of parity violation in electron\textendash{}quark scattering}},\ }\href {https://doi.org/10.1038/nature12964} {\bibfield  {journal} {\bibinfo  {journal} {Nature}\ }\textbf {\bibinfo {volume} {506}},\ \bibinfo {pages} {67} (\bibinfo {year} {2014})}\BibitemShut {NoStop}%
\bibitem [{\citenamefont {Akimov}\ \emph {et~al.}(2017)\citenamefont {Akimov} \emph {et~al.}}]{COHERENT:2017ipa}%
  \BibitemOpen
  \bibfield  {author} {\bibinfo {author} {\bibfnamefont {D.}~\bibnamefont {Akimov}} \emph {et~al.} (\bibinfo {collaboration} {COHERENT}),\ }\bibfield  {title} {\bibinfo {title} {{Observation of Coherent Elastic Neutrino-Nucleus Scattering}},\ }\href {https://doi.org/10.1126/science.aao0990} {\bibfield  {journal} {\bibinfo  {journal} {Science}\ }\textbf {\bibinfo {volume} {357}},\ \bibinfo {pages} {1123} (\bibinfo {year} {2017})},\ \Eprint {https://arxiv.org/abs/1708.01294} {arXiv:1708.01294 [nucl-ex]} \BibitemShut {NoStop}%
\bibitem [{\citenamefont {Akimov}\ \emph {et~al.}(2022)\citenamefont {Akimov} \emph {et~al.}}]{COHERENT:2021xmm}%
  \BibitemOpen
  \bibfield  {author} {\bibinfo {author} {\bibfnamefont {D.}~\bibnamefont {Akimov}} \emph {et~al.} (\bibinfo {collaboration} {COHERENT}),\ }\bibfield  {title} {\bibinfo {title} {{Measurement of the Coherent Elastic Neutrino-Nucleus Scattering Cross Section on CsI by COHERENT}},\ }\href {https://doi.org/10.1103/PhysRevLett.129.081801} {\bibfield  {journal} {\bibinfo  {journal} {Phys. Rev. Lett.}\ }\textbf {\bibinfo {volume} {129}},\ \bibinfo {pages} {081801} (\bibinfo {year} {2022})},\ \Eprint {https://arxiv.org/abs/2110.07730} {arXiv:2110.07730 [hep-ex]} \BibitemShut {NoStop}%
\bibitem [{\citenamefont {Huang}\ \emph {et~al.}(2024)\citenamefont {Huang} \emph {et~al.}}]{Huang:2024jbh}%
  \BibitemOpen
  \bibfield  {author} {\bibinfo {author} {\bibfnamefont {Y.}~\bibnamefont {Huang}} \emph {et~al.},\ }\href@noop {} {\bibinfo {title} {{Neutron radius determination of 133Cs and constraint on the weak mixing angle}}} (\bibinfo {year} {2024}),\ \Eprint {https://arxiv.org/abs/2403.03566} {arXiv:2403.03566 [nucl-ex]} \BibitemShut {NoStop}%
\bibitem [{\citenamefont {Cadeddu}\ \emph {et~al.}(2023)\citenamefont {Cadeddu}, \citenamefont {Dordei},\ and\ \citenamefont {Giunti}}]{Cadeddu:2023tkp}%
  \BibitemOpen
  \bibfield  {author} {\bibinfo {author} {\bibfnamefont {M.}~\bibnamefont {Cadeddu}}, \bibinfo {author} {\bibfnamefont {F.}~\bibnamefont {Dordei}},\ and\ \bibinfo {author} {\bibfnamefont {C.}~\bibnamefont {Giunti}},\ }\bibfield  {title} {\bibinfo {title} {{A view of coherent elastic neutrino-nucleus scattering}},\ }\href {https://doi.org/10.1209/0295-5075/ace7f0} {\bibfield  {journal} {\bibinfo  {journal} {EPL}\ }\textbf {\bibinfo {volume} {143}},\ \bibinfo {pages} {34001} (\bibinfo {year} {2023})},\ \Eprint {https://arxiv.org/abs/2307.08842} {arXiv:2307.08842 [hep-ph]} \BibitemShut {NoStop}%
\bibitem [{\citenamefont {Akimov}\ \emph {et~al.}(2020)\citenamefont {Akimov} \emph {et~al.}}]{COHERENT:2020ybo}%
  \BibitemOpen
  \bibfield  {author} {\bibinfo {author} {\bibfnamefont {D.}~\bibnamefont {Akimov}} \emph {et~al.} (\bibinfo {collaboration} {COHERENT}),\ }\href {https://doi.org/10.5281/zenodo.3903810} {\bibinfo {title} {{COHERENT Collaboration data release from the first detection of coherent elastic neutrino-nucleus scattering on argon}}} (\bibinfo {year} {2020}),\ \Eprint {https://arxiv.org/abs/2006.12659} {arXiv:2006.12659 [nucl-ex]} \BibitemShut {NoStop}%
\bibitem [{\citenamefont {Adamski}\ \emph {et~al.}(2024)\citenamefont {Adamski} \emph {et~al.}}]{Adamski:2024yqt}%
  \BibitemOpen
  \bibfield  {author} {\bibinfo {author} {\bibfnamefont {S.}~\bibnamefont {Adamski}} \emph {et~al.},\ }\href@noop {} {\bibinfo {title} {{First detection of coherent elastic neutrino-nucleus scattering on germanium}}} (\bibinfo {year} {2024}),\ \Eprint {https://arxiv.org/abs/2406.13806} {arXiv:2406.13806 [hep-ex]} \BibitemShut {NoStop}%
\bibitem [{\citenamefont {Colaresi}\ \emph {et~al.}(2022)\citenamefont {Colaresi}, \citenamefont {Collar}, \citenamefont {Hossbach}, \citenamefont {Lewis},\ and\ \citenamefont {Yocum}}]{Colaresi:2022obx}%
  \BibitemOpen
  \bibfield  {author} {\bibinfo {author} {\bibfnamefont {J.}~\bibnamefont {Colaresi}}, \bibinfo {author} {\bibfnamefont {J.~I.}\ \bibnamefont {Collar}}, \bibinfo {author} {\bibfnamefont {T.~W.}\ \bibnamefont {Hossbach}}, \bibinfo {author} {\bibfnamefont {C.~M.}\ \bibnamefont {Lewis}},\ and\ \bibinfo {author} {\bibfnamefont {K.~M.}\ \bibnamefont {Yocum}},\ }\bibfield  {title} {\bibinfo {title} {{Measurement of Coherent Elastic Neutrino-Nucleus Scattering from Reactor Antineutrinos}},\ }\href {https://doi.org/10.1103/PhysRevLett.129.211802} {\bibfield  {journal} {\bibinfo  {journal} {Phys. Rev. Lett.}\ }\textbf {\bibinfo {volume} {129}},\ \bibinfo {pages} {211802} (\bibinfo {year} {2022})},\ \Eprint {https://arxiv.org/abs/2202.09672} {arXiv:2202.09672 [hep-ex]} \BibitemShut {NoStop}%
\bibitem [{\citenamefont {Ackermann}\ \emph {et~al.}(2024)\citenamefont {Ackermann} \emph {et~al.}}]{Ackermann:2024kxo}%
  \BibitemOpen
  \bibfield  {author} {\bibinfo {author} {\bibfnamefont {N.}~\bibnamefont {Ackermann}} \emph {et~al.},\ }\href@noop {} {\bibinfo {title} {{Final CONUS results on coherent elastic neutrino nucleus scattering at the Brokdorf reactor}}} (\bibinfo {year} {2024}),\ \Eprint {https://arxiv.org/abs/2401.07684} {arXiv:2401.07684 [hep-ex]} \BibitemShut {NoStop}%
\bibitem [{\citenamefont {Collar}\ \emph {et~al.}(2021)\citenamefont {Collar}, \citenamefont {Kavner},\ and\ \citenamefont {Lewis}}]{Collar:2021fcl}%
  \BibitemOpen
  \bibfield  {author} {\bibinfo {author} {\bibfnamefont {J.~I.}\ \bibnamefont {Collar}}, \bibinfo {author} {\bibfnamefont {A.~R.~L.}\ \bibnamefont {Kavner}},\ and\ \bibinfo {author} {\bibfnamefont {C.~M.}\ \bibnamefont {Lewis}},\ }\bibfield  {title} {\bibinfo {title} {{Germanium response to sub-keV nuclear recoils: a multipronged experimental characterization}},\ }\href {https://doi.org/10.1103/PhysRevD.103.122003} {\bibfield  {journal} {\bibinfo  {journal} {Phys. Rev. D}\ }\textbf {\bibinfo {volume} {103}},\ \bibinfo {pages} {122003} (\bibinfo {year} {2021})},\ \Eprint {https://arxiv.org/abs/2102.10089} {arXiv:2102.10089 [nucl-ex]} \BibitemShut {NoStop}%
\bibitem [{\citenamefont {Atzori~Corona}\ \emph {et~al.}(2024{\natexlab{a}})\citenamefont {Atzori~Corona}, \citenamefont {Cadeddu}, \citenamefont {Cargioli}, \citenamefont {Dordei},\ and\ \citenamefont {Giunti}}]{AtzoriCorona:2023ais}%
  \BibitemOpen
  \bibfield  {author} {\bibinfo {author} {\bibfnamefont {M.}~\bibnamefont {Atzori~Corona}}, \bibinfo {author} {\bibfnamefont {M.}~\bibnamefont {Cadeddu}}, \bibinfo {author} {\bibfnamefont {N.}~\bibnamefont {Cargioli}}, \bibinfo {author} {\bibfnamefont {F.}~\bibnamefont {Dordei}},\ and\ \bibinfo {author} {\bibfnamefont {C.}~\bibnamefont {Giunti}},\ }\bibfield  {title} {\bibinfo {title} {{On the impact of the Migdal effect in reactor CE\ensuremath{\nu}NS experiments}},\ }\href {https://doi.org/10.1016/j.physletb.2024.138627} {\bibfield  {journal} {\bibinfo  {journal} {Phys. Lett. B}\ }\textbf {\bibinfo {volume} {852}},\ \bibinfo {pages} {138627} (\bibinfo {year} {2024}{\natexlab{a}})},\ \Eprint {https://arxiv.org/abs/2307.12911} {arXiv:2307.12911 [hep-ph]} \BibitemShut {NoStop}%
\bibitem [{\citenamefont {Atzori~Corona}\ \emph {et~al.}(2024{\natexlab{b}})\citenamefont {Atzori~Corona}, \citenamefont {Cadeddu}, \citenamefont {Cargioli}, \citenamefont {Dordei},\ and\ \citenamefont {Giunti}}]{AtzoriCorona:2024rtv}%
  \BibitemOpen
  \bibfield  {author} {\bibinfo {author} {\bibfnamefont {M.}~\bibnamefont {Atzori~Corona}}, \bibinfo {author} {\bibfnamefont {M.}~\bibnamefont {Cadeddu}}, \bibinfo {author} {\bibfnamefont {N.}~\bibnamefont {Cargioli}}, \bibinfo {author} {\bibfnamefont {F.}~\bibnamefont {Dordei}},\ and\ \bibinfo {author} {\bibfnamefont {C.}~\bibnamefont {Giunti}},\ }\bibfield  {title} {\bibinfo {title} {{Momentum dependent flavor radiative corrections to the coherent elastic neutrino-nucleus scattering for the neutrino charge-radius determination}},\ }\href {https://doi.org/10.1007/JHEP05(2024)271} {\bibfield  {journal} {\bibinfo  {journal} {JHEP}\ }\textbf {\bibinfo {volume} {05}},\ \bibinfo {pages} {271}},\ \Eprint {https://arxiv.org/abs/2402.16709} {arXiv:2402.16709 [hep-ph]} \BibitemShut {NoStop}%
\bibitem [{\citenamefont {Atzori~Corona}\ \emph {et~al.}(2023)\citenamefont {Atzori~Corona}, \citenamefont {Cadeddu}, \citenamefont {Cargioli}, \citenamefont {Dordei}, \citenamefont {Giunti},\ and\ \citenamefont {Masia}}]{AtzoriCorona:2023ktl}%
  \BibitemOpen
  \bibfield  {author} {\bibinfo {author} {\bibfnamefont {M.}~\bibnamefont {Atzori~Corona}}, \bibinfo {author} {\bibfnamefont {M.}~\bibnamefont {Cadeddu}}, \bibinfo {author} {\bibfnamefont {N.}~\bibnamefont {Cargioli}}, \bibinfo {author} {\bibfnamefont {F.}~\bibnamefont {Dordei}}, \bibinfo {author} {\bibfnamefont {C.}~\bibnamefont {Giunti}},\ and\ \bibinfo {author} {\bibfnamefont {G.}~\bibnamefont {Masia}},\ }\bibfield  {title} {\bibinfo {title} {{Nuclear neutron radius and weak mixing angle measurements from latest COHERENT CsI and atomic parity violation Cs data}},\ }\href {https://doi.org/10.1140/epjc/s10052-023-11849-5} {\bibfield  {journal} {\bibinfo  {journal} {Eur. Phys. J. C}\ }\textbf {\bibinfo {volume} {83}},\ \bibinfo {pages} {683} (\bibinfo {year} {2023})},\ \Eprint {https://arxiv.org/abs/2303.09360} {arXiv:2303.09360 [nucl-ex]} \BibitemShut {NoStop}%
\bibitem [{\citenamefont {Atzori~Corona}\ \emph {et~al.}(2022)\citenamefont {Atzori~Corona}, \citenamefont {Cadeddu}, \citenamefont {Cargioli}, \citenamefont {Dordei}, \citenamefont {Giunti}, \citenamefont {Li}, \citenamefont {Ternes},\ and\ \citenamefont {Zhang}}]{AtzoriCorona:2022qrf}%
  \BibitemOpen
  \bibfield  {author} {\bibinfo {author} {\bibfnamefont {M.}~\bibnamefont {Atzori~Corona}}, \bibinfo {author} {\bibfnamefont {M.}~\bibnamefont {Cadeddu}}, \bibinfo {author} {\bibfnamefont {N.}~\bibnamefont {Cargioli}}, \bibinfo {author} {\bibfnamefont {F.}~\bibnamefont {Dordei}}, \bibinfo {author} {\bibfnamefont {C.}~\bibnamefont {Giunti}}, \bibinfo {author} {\bibfnamefont {Y.~F.}\ \bibnamefont {Li}}, \bibinfo {author} {\bibfnamefont {C.~A.}\ \bibnamefont {Ternes}},\ and\ \bibinfo {author} {\bibfnamefont {Y.~Y.}\ \bibnamefont {Zhang}},\ }\bibfield  {title} {\bibinfo {title} {{Impact of the Dresden-II and COHERENT neutrino scattering data on neutrino electromagnetic properties and electroweak physics}},\ }\href {https://doi.org/10.1007/JHEP09(2022)164} {\bibfield  {journal} {\bibinfo  {journal} {JHEP}\ }\textbf {\bibinfo {volume} {09}},\ \bibinfo {pages} {164}},\ \Eprint {https://arxiv.org/abs/2205.09484} {arXiv:2205.09484 [hep-ph]} \BibitemShut {NoStop}%
\bibitem [{\citenamefont {Cadeddu}\ \emph {et~al.}(2021{\natexlab{b}})\citenamefont {Cadeddu}, \citenamefont {Cargioli}, \citenamefont {Dordei}, \citenamefont {Giunti}, \citenamefont {Li}, \citenamefont {Picciau}, \citenamefont {Ternes},\ and\ \citenamefont {Zhang}}]{Cadeddu:2021ijh}%
  \BibitemOpen
  \bibfield  {author} {\bibinfo {author} {\bibfnamefont {M.}~\bibnamefont {Cadeddu}}, \bibinfo {author} {\bibfnamefont {N.}~\bibnamefont {Cargioli}}, \bibinfo {author} {\bibfnamefont {F.}~\bibnamefont {Dordei}}, \bibinfo {author} {\bibfnamefont {C.}~\bibnamefont {Giunti}}, \bibinfo {author} {\bibfnamefont {Y.~F.}\ \bibnamefont {Li}}, \bibinfo {author} {\bibfnamefont {E.}~\bibnamefont {Picciau}}, \bibinfo {author} {\bibfnamefont {C.~A.}\ \bibnamefont {Ternes}},\ and\ \bibinfo {author} {\bibfnamefont {Y.~Y.}\ \bibnamefont {Zhang}},\ }\bibfield  {title} {\bibinfo {title} {{New insights into nuclear physics and weak mixing angle using electroweak probes}},\ }\href {https://doi.org/10.1103/PhysRevC.104.065502} {\bibfield  {journal} {\bibinfo  {journal} {Phys. Rev. C}\ }\textbf {\bibinfo {volume} {104}},\ \bibinfo {pages} {065502} (\bibinfo {year} {2021}{\natexlab{b}})},\ \Eprint {https://arxiv.org/abs/2102.06153} {arXiv:2102.06153 [hep-ph]} \BibitemShut {NoStop}%
\bibitem [{\citenamefont {Cadeddu}\ \emph {et~al.}(2020{\natexlab{a}})\citenamefont {Cadeddu}, \citenamefont {Dordei}, \citenamefont {Giunti}, \citenamefont {Li},\ and\ \citenamefont {Zhang}}]{Cadeddu:2019eta}%
  \BibitemOpen
  \bibfield  {author} {\bibinfo {author} {\bibfnamefont {M.}~\bibnamefont {Cadeddu}}, \bibinfo {author} {\bibfnamefont {F.}~\bibnamefont {Dordei}}, \bibinfo {author} {\bibfnamefont {C.}~\bibnamefont {Giunti}}, \bibinfo {author} {\bibfnamefont {Y.~F.}\ \bibnamefont {Li}},\ and\ \bibinfo {author} {\bibfnamefont {Y.~Y.}\ \bibnamefont {Zhang}},\ }\bibfield  {title} {\bibinfo {title} {{Neutrino, electroweak, and nuclear physics from COHERENT elastic neutrino-nucleus scattering with refined quenching factor}},\ }\href {https://doi.org/10.1103/PhysRevD.101.033004} {\bibfield  {journal} {\bibinfo  {journal} {Phys. Rev.}\ }\textbf {\bibinfo {volume} {D101}},\ \bibinfo {pages} {033004} (\bibinfo {year} {2020}{\natexlab{a}})},\ \Eprint {https://arxiv.org/abs/1908.06045} {arXiv:1908.06045 [hep-ph]} \BibitemShut {NoStop}%
\bibitem [{\citenamefont {Miranda}\ \emph {et~al.}(2020)\citenamefont {Miranda}, \citenamefont {Papoulias}, \citenamefont {Garcia}, \citenamefont {Sanders}, \citenamefont {Tortola},\ and\ \citenamefont {Valle}}]{Miranda:2020tif}%
  \BibitemOpen
  \bibfield  {author} {\bibinfo {author} {\bibfnamefont {O.~G.}\ \bibnamefont {Miranda}}, \bibinfo {author} {\bibfnamefont {D.~K.}\ \bibnamefont {Papoulias}}, \bibinfo {author} {\bibfnamefont {G.~S.}\ \bibnamefont {Garcia}}, \bibinfo {author} {\bibfnamefont {O.}~\bibnamefont {Sanders}}, \bibinfo {author} {\bibfnamefont {M.}~\bibnamefont {Tortola}},\ and\ \bibinfo {author} {\bibfnamefont {J.~W.~F.}\ \bibnamefont {Valle}},\ }\bibfield  {title} {\bibinfo {title} {{Implications of the first detection of coherent elastic neutrino-nucleus scattering (CEvNS) with Liquid Argon}},\ }\href@noop {} {\bibfield  {journal} {\bibinfo  {journal} {JHEP}\ }\textbf {\bibinfo {volume} {2005}},\ \bibinfo {pages} {130}},\ \Eprint {https://arxiv.org/abs/arXiv:2003.12050} {arXiv:2003.12050 [hep-ph]} \BibitemShut {NoStop}%
\bibitem [{\citenamefont {Papoulias}\ \emph {et~al.}(2020)\citenamefont {Papoulias}, \citenamefont {Kosmas}, \citenamefont {Sahu}, \citenamefont {Kota},\ and\ \citenamefont {Hota}}]{Papoulias:2019lfi}%
  \BibitemOpen
  \bibfield  {author} {\bibinfo {author} {\bibfnamefont {D.~K.}\ \bibnamefont {Papoulias}}, \bibinfo {author} {\bibfnamefont {T.~S.}\ \bibnamefont {Kosmas}}, \bibinfo {author} {\bibfnamefont {R.}~\bibnamefont {Sahu}}, \bibinfo {author} {\bibfnamefont {V.~K.~B.}\ \bibnamefont {Kota}},\ and\ \bibinfo {author} {\bibfnamefont {M.}~\bibnamefont {Hota}},\ }\bibfield  {title} {\bibinfo {title} {{Constraining nuclear physics parameters with current and future COHERENT data}},\ }\href@noop {} {\bibfield  {journal} {\bibinfo  {journal} {Phys.Lett.}\ }\textbf {\bibinfo {volume} {B800}},\ \bibinfo {pages} {135133} (\bibinfo {year} {2020})},\ \Eprint {https://arxiv.org/abs/arXiv:1903.03722} {arXiv:1903.03722 [hep-ph]} \BibitemShut {NoStop}%
\bibitem [{\citenamefont {De~Romeri}\ \emph {et~al.}(2023)\citenamefont {De~Romeri}, \citenamefont {Miranda}, \citenamefont {Papoulias}, \citenamefont {Sanchez~Garcia}, \citenamefont {Tortola},\ and\ \citenamefont {Valle}}]{DeRomeri:2022twg}%
  \BibitemOpen
  \bibfield  {author} {\bibinfo {author} {\bibfnamefont {V.}~\bibnamefont {De~Romeri}}, \bibinfo {author} {\bibfnamefont {O.~G.}\ \bibnamefont {Miranda}}, \bibinfo {author} {\bibfnamefont {D.~K.}\ \bibnamefont {Papoulias}}, \bibinfo {author} {\bibfnamefont {G.}~\bibnamefont {Sanchez~Garcia}}, \bibinfo {author} {\bibfnamefont {M.}~\bibnamefont {Tortola}},\ and\ \bibinfo {author} {\bibfnamefont {J.~W.~F.}\ \bibnamefont {Valle}},\ }\bibfield  {title} {\bibinfo {title} {{Physics implications of a combined analysis of COHERENT CsI and LAr data}},\ }\href@noop {} {\bibfield  {journal} {\bibinfo  {journal} {JHEP}\ }\textbf {\bibinfo {volume} {04}},\ \bibinfo {pages} {035}},\ \Eprint {https://arxiv.org/abs/arXiv:2211.11905} {arXiv:2211.11905 [hep-ph]} \BibitemShut {NoStop}%
\bibitem [{\citenamefont {Corona}\ \emph {et~al.}(2022)\citenamefont {Corona}, \citenamefont {Cadeddu}, \citenamefont {Cargioli}, \citenamefont {Finelli},\ and\ \citenamefont {Vorabbi}}]{Corona:2021yfd}%
  \BibitemOpen
  \bibfield  {author} {\bibinfo {author} {\bibfnamefont {M.~A.}\ \bibnamefont {Corona}}, \bibinfo {author} {\bibfnamefont {M.}~\bibnamefont {Cadeddu}}, \bibinfo {author} {\bibfnamefont {N.}~\bibnamefont {Cargioli}}, \bibinfo {author} {\bibfnamefont {P.}~\bibnamefont {Finelli}},\ and\ \bibinfo {author} {\bibfnamefont {M.}~\bibnamefont {Vorabbi}},\ }\bibfield  {title} {\bibinfo {title} {{Incorporating the weak mixing angle dependence to reconcile the neutron skin measurement on Pb208 by PREX-II}},\ }\href {https://doi.org/10.1103/PhysRevC.105.055503} {\bibfield  {journal} {\bibinfo  {journal} {Phys. Rev. C}\ }\textbf {\bibinfo {volume} {105}},\ \bibinfo {pages} {055503} (\bibinfo {year} {2022})},\ \Eprint {https://arxiv.org/abs/2112.09717} {arXiv:2112.09717 [hep-ph]} \BibitemShut {NoStop}%
\bibitem [{\citenamefont {Adhikari}\ \emph {et~al.}(2021)\citenamefont {Adhikari} \emph {et~al.}}]{PREX:2021umo}%
  \BibitemOpen
  \bibfield  {author} {\bibinfo {author} {\bibfnamefont {D.}~\bibnamefont {Adhikari}} \emph {et~al.} (\bibinfo {collaboration} {PREX}),\ }\bibfield  {title} {\bibinfo {title} {{Accurate Determination of the Neutron Skin Thickness of $^{208}$Pb through Parity-Violation in Electron Scattering}},\ }\href {https://doi.org/10.1103/PhysRevLett.126.172502} {\bibfield  {journal} {\bibinfo  {journal} {Phys. Rev. Lett.}\ }\textbf {\bibinfo {volume} {126}},\ \bibinfo {pages} {172502} (\bibinfo {year} {2021})},\ \Eprint {https://arxiv.org/abs/2102.10767} {arXiv:2102.10767 [nucl-ex]} \BibitemShut {NoStop}%
\bibitem [{\citenamefont {Thiel}\ \emph {et~al.}(2019)\citenamefont {Thiel}, \citenamefont {Sfienti}, \citenamefont {Piekarewicz}, \citenamefont {Horowitz},\ and\ \citenamefont {Vanderhaeghen}}]{Thiel:2019tkm}%
  \BibitemOpen
  \bibfield  {author} {\bibinfo {author} {\bibfnamefont {M.}~\bibnamefont {Thiel}}, \bibinfo {author} {\bibfnamefont {C.}~\bibnamefont {Sfienti}}, \bibinfo {author} {\bibfnamefont {J.}~\bibnamefont {Piekarewicz}}, \bibinfo {author} {\bibfnamefont {C.~J.}\ \bibnamefont {Horowitz}},\ and\ \bibinfo {author} {\bibfnamefont {M.}~\bibnamefont {Vanderhaeghen}},\ }\bibfield  {title} {\bibinfo {title} {{Neutron skins of atomic nuclei: per aspera ad astra}},\ }\href {https://doi.org/10.1088/1361-6471/ab2c6d} {\bibfield  {journal} {\bibinfo  {journal} {J. Phys. G}\ }\textbf {\bibinfo {volume} {46}},\ \bibinfo {pages} {093003} (\bibinfo {year} {2019})},\ \Eprint {https://arxiv.org/abs/1904.12269} {arXiv:1904.12269 [nucl-ex]} \BibitemShut {NoStop}%
\bibitem [{\citenamefont {Angeli}\ and\ \citenamefont {Marinova}(2013)}]{Angeli:2013epw}%
  \BibitemOpen
  \bibfield  {author} {\bibinfo {author} {\bibfnamefont {I.}~\bibnamefont {Angeli}}\ and\ \bibinfo {author} {\bibfnamefont {K.~P.}\ \bibnamefont {Marinova}},\ }\bibfield  {title} {\bibinfo {title} {{Table of experimental nuclear ground state charge radii: An update}},\ }\href {https://doi.org/10.1016/j.adt.2011.12.006} {\bibfield  {journal} {\bibinfo  {journal} {Atom. Data Nucl. Data Tabl.}\ }\textbf {\bibinfo {volume} {99}},\ \bibinfo {pages} {69} (\bibinfo {year} {2013})}\BibitemShut {NoStop}%
\bibitem [{\citenamefont {Trzcinska}\ \emph {et~al.}(2001)\citenamefont {Trzcinska}, \citenamefont {Jastrzebski}, \citenamefont {Lubinski}, \citenamefont {Hartmann}, \citenamefont {Schmidt}, \citenamefont {von Egidy},\ and\ \citenamefont {Klos}}]{Trzcinska:2001sy}%
  \BibitemOpen
  \bibfield  {author} {\bibinfo {author} {\bibfnamefont {A.}~\bibnamefont {Trzcinska}}, \bibinfo {author} {\bibfnamefont {J.}~\bibnamefont {Jastrzebski}}, \bibinfo {author} {\bibfnamefont {P.}~\bibnamefont {Lubinski}}, \bibinfo {author} {\bibfnamefont {F.~J.}\ \bibnamefont {Hartmann}}, \bibinfo {author} {\bibfnamefont {R.}~\bibnamefont {Schmidt}}, \bibinfo {author} {\bibfnamefont {T.}~\bibnamefont {von Egidy}},\ and\ \bibinfo {author} {\bibfnamefont {B.}~\bibnamefont {Klos}},\ }\bibfield  {title} {\bibinfo {title} {{Neutron density distributions deduced from anti-protonic atoms}},\ }\href {https://doi.org/10.1103/PhysRevLett.87.082501} {\bibfield  {journal} {\bibinfo  {journal} {Phys. Rev. Lett.}\ }\textbf {\bibinfo {volume} {87}},\ \bibinfo {pages} {082501} (\bibinfo {year} {2001})}\BibitemShut {NoStop}%
\bibitem [{\citenamefont {Sahoo}\ \emph {et~al.}(2021)\citenamefont {Sahoo}, \citenamefont {Das},\ and\ \citenamefont {Spiesberger}}]{Sahoo:2021thl}%
  \BibitemOpen
  \bibfield  {author} {\bibinfo {author} {\bibfnamefont {B.~K.}\ \bibnamefont {Sahoo}}, \bibinfo {author} {\bibfnamefont {B.~P.}\ \bibnamefont {Das}},\ and\ \bibinfo {author} {\bibfnamefont {H.}~\bibnamefont {Spiesberger}},\ }\bibfield  {title} {\bibinfo {title} {{New physics constraints from atomic parity violation in Cs133}},\ }\href {https://doi.org/10.1103/PhysRevD.103.L111303} {\bibfield  {journal} {\bibinfo  {journal} {Phys. Rev. D}\ }\textbf {\bibinfo {volume} {103}},\ \bibinfo {pages} {L111303} (\bibinfo {year} {2021})},\ \Eprint {https://arxiv.org/abs/2101.10095} {arXiv:2101.10095 [hep-ph]} \BibitemShut {NoStop}%
\bibitem [{\citenamefont {Tran~Tan}\ \emph {et~al.}(2022)\citenamefont {Tran~Tan}, \citenamefont {Xiao},\ and\ \citenamefont {Derevianko}}]{Tran_Tan_2022}%
  \BibitemOpen
  \bibfield  {author} {\bibinfo {author} {\bibfnamefont {H.~B.}\ \bibnamefont {Tran~Tan}}, \bibinfo {author} {\bibfnamefont {D.}~\bibnamefont {Xiao}},\ and\ \bibinfo {author} {\bibfnamefont {A.}~\bibnamefont {Derevianko}},\ }\bibfield  {title} {\bibinfo {title} {Parity-mixed coupled-cluster formalism for computing parity-violating amplitudes},\ }\bibfield  {journal} {\bibinfo  {journal} {Physical Review A}\ }\textbf {\bibinfo {volume} {105}},\ \href {https://doi.org/10.1103/physreva.105.022803} {10.1103/physreva.105.022803} (\bibinfo {year} {2022})\BibitemShut {NoStop}%
\bibitem [{\citenamefont {Porsev}\ \emph {et~al.}(2010)\citenamefont {Porsev}, \citenamefont {Beloy},\ and\ \citenamefont {Derevianko}}]{Porsev:2010de}%
  \BibitemOpen
  \bibfield  {author} {\bibinfo {author} {\bibfnamefont {S.~G.}\ \bibnamefont {Porsev}}, \bibinfo {author} {\bibfnamefont {K.}~\bibnamefont {Beloy}},\ and\ \bibinfo {author} {\bibfnamefont {A.}~\bibnamefont {Derevianko}},\ }\bibfield  {title} {\bibinfo {title} {{Precision determination of weak charge of $^{133}$Cs from atomic parity violation}},\ }\href {https://doi.org/10.1103/PhysRevD.82.036008} {\bibfield  {journal} {\bibinfo  {journal} {Phys. Rev. D}\ }\textbf {\bibinfo {volume} {82}},\ \bibinfo {pages} {036008} (\bibinfo {year} {2010})},\ \Eprint {https://arxiv.org/abs/1006.4193} {arXiv:1006.4193 [hep-ph]} \BibitemShut {NoStop}%
\bibitem [{\citenamefont {Cadeddu}\ and\ \citenamefont {Dordei}(2019)}]{Cadeddu:2018izq}%
  \BibitemOpen
  \bibfield  {author} {\bibinfo {author} {\bibfnamefont {M.}~\bibnamefont {Cadeddu}}\ and\ \bibinfo {author} {\bibfnamefont {F.}~\bibnamefont {Dordei}},\ }\bibfield  {title} {\bibinfo {title} {{Reinterpreting the weak mixing angle from atomic parity violation in view of the Cs neutron rms radius measurement from COHERENT}},\ }\href@noop {} {\bibfield  {journal} {\bibinfo  {journal} {Phys.Rev.}\ }\textbf {\bibinfo {volume} {D99}},\ \bibinfo {pages} {033010} (\bibinfo {year} {2019})},\ \Eprint {https://arxiv.org/abs/arXiv:1808.10202} {arXiv:1808.10202 [hep-ph]} \BibitemShut {NoStop}%
\bibitem [{\citenamefont {Fricke}\ \emph {et~al.}(1995)\citenamefont {Fricke}, \citenamefont {Bernhardt}, \citenamefont {Heilig}, \citenamefont {Schaller}, \citenamefont {Schellenberg}, \citenamefont {Shera},\ and\ \citenamefont {de~Jager}}]{Fricke:1995zz}%
  \BibitemOpen
  \bibfield  {author} {\bibinfo {author} {\bibfnamefont {G.}~\bibnamefont {Fricke}}, \bibinfo {author} {\bibfnamefont {C.}~\bibnamefont {Bernhardt}}, \bibinfo {author} {\bibfnamefont {K.}~\bibnamefont {Heilig}}, \bibinfo {author} {\bibfnamefont {L.~A.}\ \bibnamefont {Schaller}}, \bibinfo {author} {\bibfnamefont {L.}~\bibnamefont {Schellenberg}}, \bibinfo {author} {\bibfnamefont {E.~B.}\ \bibnamefont {Shera}},\ and\ \bibinfo {author} {\bibfnamefont {C.~W.}\ \bibnamefont {de~Jager}},\ }\bibfield  {title} {\bibinfo {title} {{Nuclear Ground State Charge Radii from Electromagnetic Interactions}},\ }\href {https://doi.org/10.1006/adnd.1995.1007} {\bibfield  {journal} {\bibinfo  {journal} {Atom. Data Nucl. Data Tabl.}\ }\textbf {\bibinfo {volume} {60}},\ \bibinfo {pages} {177} (\bibinfo {year} {1995})}\BibitemShut {NoStop}%
\bibitem [{\citenamefont {Cadeddu}\ \emph {et~al.}(2020{\natexlab{b}})\citenamefont {Cadeddu}, \citenamefont {Dordei}, \citenamefont {Giunti}, \citenamefont {Li}, \citenamefont {Picciau},\ and\ \citenamefont {Zhang}}]{Cadeddu:2020lky}%
  \BibitemOpen
  \bibfield  {author} {\bibinfo {author} {\bibfnamefont {M.}~\bibnamefont {Cadeddu}}, \bibinfo {author} {\bibfnamefont {F.}~\bibnamefont {Dordei}}, \bibinfo {author} {\bibfnamefont {C.}~\bibnamefont {Giunti}}, \bibinfo {author} {\bibfnamefont {Y.~F.}\ \bibnamefont {Li}}, \bibinfo {author} {\bibfnamefont {E.}~\bibnamefont {Picciau}},\ and\ \bibinfo {author} {\bibfnamefont {Y.~Y.}\ \bibnamefont {Zhang}},\ }\bibfield  {title} {\bibinfo {title} {{Physics results from the first COHERENT observation of coherent elastic neutrino-nucleus scattering in argon and their combination with cesium-iodide data}},\ }\href {https://doi.org/10.1103/PhysRevD.102.015030} {\bibfield  {journal} {\bibinfo  {journal} {Phys. Rev. D}\ }\textbf {\bibinfo {volume} {102}},\ \bibinfo {pages} {015030} (\bibinfo {year} {2020}{\natexlab{b}})},\ \Eprint {https://arxiv.org/abs/2005.01645} {arXiv:2005.01645 [hep-ph]} \BibitemShut {NoStop}%
\bibitem [{\citenamefont {Lattimer}(2023{\natexlab{a}})}]{Lattimer:2023rpe}%
  \BibitemOpen
  \bibfield  {author} {\bibinfo {author} {\bibfnamefont {J.~M.}\ \bibnamefont {Lattimer}},\ }\bibfield  {title} {\bibinfo {title} {{Constraints on Nuclear Symmetry Energy Parameters}},\ }\href {https://doi.org/10.3390/particles6010003} {\bibfield  {journal} {\bibinfo  {journal} {Particles}\ }\textbf {\bibinfo {volume} {6}},\ \bibinfo {pages} {30} (\bibinfo {year} {2023}{\natexlab{a}})},\ \Eprint {https://arxiv.org/abs/2301.03666} {arXiv:2301.03666 [nucl-th]} \BibitemShut {NoStop}%
\bibitem [{\citenamefont {Zhang}\ \emph {et~al.}(2021{\natexlab{a}})\citenamefont {Zhang}, \citenamefont {Tu}, \citenamefont {Sarriguren}, \citenamefont {Yue}, \citenamefont {Zeng}, \citenamefont {Sun}, \citenamefont {Wang}, \citenamefont {Zhang}, \citenamefont {Zhou},\ and\ \citenamefont {Litvinov}}]{PhysRevC.104.034303}%
  \BibitemOpen
  \bibfield  {author} {\bibinfo {author} {\bibfnamefont {J.~T.}\ \bibnamefont {Zhang}}, \bibinfo {author} {\bibfnamefont {X.~L.}\ \bibnamefont {Tu}}, \bibinfo {author} {\bibfnamefont {P.}~\bibnamefont {Sarriguren}}, \bibinfo {author} {\bibfnamefont {K.}~\bibnamefont {Yue}}, \bibinfo {author} {\bibfnamefont {Q.}~\bibnamefont {Zeng}}, \bibinfo {author} {\bibfnamefont {Z.~Y.}\ \bibnamefont {Sun}}, \bibinfo {author} {\bibfnamefont {M.}~\bibnamefont {Wang}}, \bibinfo {author} {\bibfnamefont {Y.~H.}\ \bibnamefont {Zhang}}, \bibinfo {author} {\bibfnamefont {X.~H.}\ \bibnamefont {Zhou}},\ and\ \bibinfo {author} {\bibfnamefont {Y.~A.}\ \bibnamefont {Litvinov}},\ }\bibfield  {title} {\bibinfo {title} {Systematic trends of neutron skin thickness versus relative neutron excess},\ }\href {https://doi.org/10.1103/PhysRevC.104.034303} {\bibfield  {journal} {\bibinfo  {journal} {Phys. Rev. C}\ }\textbf {\bibinfo {volume} {104}},\ \bibinfo {pages} {034303} (\bibinfo {year} {2021}{\natexlab{a}})}\BibitemShut {NoStop}%
\bibitem [{\citenamefont {Giacalone}\ \emph {et~al.}(2023{\natexlab{a}})\citenamefont {Giacalone}, \citenamefont {Nijs},\ and\ \citenamefont {van~der Schee}}]{PhysRevLett.131.202302}%
  \BibitemOpen
  \bibfield  {author} {\bibinfo {author} {\bibfnamefont {G.}~\bibnamefont {Giacalone}}, \bibinfo {author} {\bibfnamefont {G.}~\bibnamefont {Nijs}},\ and\ \bibinfo {author} {\bibfnamefont {W.}~\bibnamefont {van~der Schee}},\ }\bibfield  {title} {\bibinfo {title} {Determination of the neutron skin of $^{208}\mathrm{Pb}$ from ultrarelativistic nuclear collisions},\ }\href {https://doi.org/10.1103/PhysRevLett.131.202302} {\bibfield  {journal} {\bibinfo  {journal} {Phys. Rev. Lett.}\ }\textbf {\bibinfo {volume} {131}},\ \bibinfo {pages} {202302} (\bibinfo {year} {2023}{\natexlab{a}})}\BibitemShut {NoStop}%
\bibitem [{\citenamefont {Giacalone}\ \emph {et~al.}(2023{\natexlab{b}})\citenamefont {Giacalone}, \citenamefont {Nijs},\ and\ \citenamefont {van~der Schee}}]{Giacalone:2023cet}%
  \BibitemOpen
  \bibfield  {author} {\bibinfo {author} {\bibfnamefont {G.}~\bibnamefont {Giacalone}}, \bibinfo {author} {\bibfnamefont {G.}~\bibnamefont {Nijs}},\ and\ \bibinfo {author} {\bibfnamefont {W.}~\bibnamefont {van~der Schee}},\ }\bibfield  {title} {\bibinfo {title} {{Determination of the Neutron Skin of Pb208 from Ultrarelativistic Nuclear Collisions}},\ }\href {https://doi.org/10.1103/PhysRevLett.131.202302} {\bibfield  {journal} {\bibinfo  {journal} {Phys. Rev. Lett.}\ }\textbf {\bibinfo {volume} {131}},\ \bibinfo {pages} {202302} (\bibinfo {year} {2023}{\natexlab{b}})},\ \Eprint {https://arxiv.org/abs/2305.00015} {arXiv:2305.00015 [nucl-th]} \BibitemShut {NoStop}%
\bibitem [{\citenamefont {Zhang}\ \emph {et~al.}(2021{\natexlab{b}})\citenamefont {Zhang}, \citenamefont {Tu}, \citenamefont {Sarriguren}, \citenamefont {Yue}, \citenamefont {Zeng}, \citenamefont {Sun}, \citenamefont {Wang}, \citenamefont {Zhang}, \citenamefont {Zhou},\ and\ \citenamefont {Litvinov}}]{Zhang:2021jwh}%
  \BibitemOpen
  \bibfield  {author} {\bibinfo {author} {\bibfnamefont {J.~T.}\ \bibnamefont {Zhang}}, \bibinfo {author} {\bibfnamefont {X.~L.}\ \bibnamefont {Tu}}, \bibinfo {author} {\bibfnamefont {P.}~\bibnamefont {Sarriguren}}, \bibinfo {author} {\bibfnamefont {K.}~\bibnamefont {Yue}}, \bibinfo {author} {\bibfnamefont {Q.}~\bibnamefont {Zeng}}, \bibinfo {author} {\bibfnamefont {Z.~Y.}\ \bibnamefont {Sun}}, \bibinfo {author} {\bibfnamefont {M.}~\bibnamefont {Wang}}, \bibinfo {author} {\bibfnamefont {Y.~H.}\ \bibnamefont {Zhang}}, \bibinfo {author} {\bibfnamefont {X.~H.}\ \bibnamefont {Zhou}},\ and\ \bibinfo {author} {\bibfnamefont {Y.~A.}\ \bibnamefont {Litvinov}},\ }\bibfield  {title} {\bibinfo {title} {{Systematic trends of neutron skin thickness versus relative neutron excess}},\ }\href {https://doi.org/10.1103/PhysRevC.104.034303} {\bibfield  {journal} {\bibinfo  {journal} {Phys. Rev. C}\ }\textbf {\bibinfo {volume} {104}},\ \bibinfo {pages} {034303} (\bibinfo {year} {2021}{\natexlab{b}})},\ \Eprint
  {https://arxiv.org/abs/2109.03417} {arXiv:2109.03417 [nucl-th]} \BibitemShut {NoStop}%
\bibitem [{\citenamefont {Lattimer}(2023{\natexlab{b}})}]{Lattimer:2023xjm}%
  \BibitemOpen
  \bibfield  {author} {\bibinfo {author} {\bibfnamefont {J.~M.}\ \bibnamefont {Lattimer}},\ }\bibfield  {title} {\bibinfo {title} {{Constraints on the Nuclear Symmetry Energy from Experiments, Theory and Observations}},\ }\href {https://doi.org/10.1088/1742-6596/2536/1/012009} {\bibfield  {journal} {\bibinfo  {journal} {J. Phys. Conf. Ser.}\ }\textbf {\bibinfo {volume} {2536}},\ \bibinfo {pages} {012009} (\bibinfo {year} {2023}{\natexlab{b}})},\ \Eprint {https://arxiv.org/abs/2308.08001} {arXiv:2308.08001 [nucl-th]} \BibitemShut {NoStop}%
\bibitem [{\citenamefont {Piekarewicz}\ \emph {et~al.}(2012)\citenamefont {Piekarewicz}, \citenamefont {Agrawal}, \citenamefont {Col\`o}, \citenamefont {Nazarewicz}, \citenamefont {Paar}, \citenamefont {Reinhard}, \citenamefont {Roca-Maza},\ and\ \citenamefont {Vretenar}}]{PhysRevC.85.041302}%
  \BibitemOpen
  \bibfield  {author} {\bibinfo {author} {\bibfnamefont {J.}~\bibnamefont {Piekarewicz}}, \bibinfo {author} {\bibfnamefont {B.~K.}\ \bibnamefont {Agrawal}}, \bibinfo {author} {\bibfnamefont {G.}~\bibnamefont {Col\`o}}, \bibinfo {author} {\bibfnamefont {W.}~\bibnamefont {Nazarewicz}}, \bibinfo {author} {\bibfnamefont {N.}~\bibnamefont {Paar}}, \bibinfo {author} {\bibfnamefont {P.-G.}\ \bibnamefont {Reinhard}}, \bibinfo {author} {\bibfnamefont {X.}~\bibnamefont {Roca-Maza}},\ and\ \bibinfo {author} {\bibfnamefont {D.}~\bibnamefont {Vretenar}},\ }\bibfield  {title} {\bibinfo {title} {Electric dipole polarizability and the neutron skin},\ }\href {https://doi.org/10.1103/PhysRevC.85.041302} {\bibfield  {journal} {\bibinfo  {journal} {Phys. Rev. C}\ }\textbf {\bibinfo {volume} {85}},\ \bibinfo {pages} {041302} (\bibinfo {year} {2012})}\BibitemShut {NoStop}%
\bibitem [{\citenamefont {Roca-Maza}\ \emph {et~al.}(2015)\citenamefont {Roca-Maza}, \citenamefont {Vi\~nas}, \citenamefont {Centelles}, \citenamefont {Agrawal}, \citenamefont {Col\`o}, \citenamefont {Paar}, \citenamefont {Piekarewicz},\ and\ \citenamefont {Vretenar}}]{PhysRevC.92.064304}%
  \BibitemOpen
  \bibfield  {author} {\bibinfo {author} {\bibfnamefont {X.}~\bibnamefont {Roca-Maza}}, \bibinfo {author} {\bibfnamefont {X.}~\bibnamefont {Vi\~nas}}, \bibinfo {author} {\bibfnamefont {M.}~\bibnamefont {Centelles}}, \bibinfo {author} {\bibfnamefont {B.~K.}\ \bibnamefont {Agrawal}}, \bibinfo {author} {\bibfnamefont {G.}~\bibnamefont {Col\`o}}, \bibinfo {author} {\bibfnamefont {N.}~\bibnamefont {Paar}}, \bibinfo {author} {\bibfnamefont {J.}~\bibnamefont {Piekarewicz}},\ and\ \bibinfo {author} {\bibfnamefont {D.}~\bibnamefont {Vretenar}},\ }\bibfield  {title} {\bibinfo {title} {Neutron skin thickness from the measured electric dipole polarizability in $^{68}\text{Ni}$, $^{120}\text{Sn}$, and $^{208}\text{Pb}$},\ }\href {https://doi.org/10.1103/PhysRevC.92.064304} {\bibfield  {journal} {\bibinfo  {journal} {Phys. Rev. C}\ }\textbf {\bibinfo {volume} {92}},\ \bibinfo {pages} {064304} (\bibinfo {year} {2015})}\BibitemShut {NoStop}%
\bibitem [{\citenamefont {Piekarewicz}(2021)}]{PhysRevC.104.024329}%
  \BibitemOpen
  \bibfield  {author} {\bibinfo {author} {\bibfnamefont {J.}~\bibnamefont {Piekarewicz}},\ }\bibfield  {title} {\bibinfo {title} {Implications of prex-2 on the electric dipole polarizability of neutron-rich nuclei},\ }\href {https://doi.org/10.1103/PhysRevC.104.024329} {\bibfield  {journal} {\bibinfo  {journal} {Phys. Rev. C}\ }\textbf {\bibinfo {volume} {104}},\ \bibinfo {pages} {024329} (\bibinfo {year} {2021})}\BibitemShut {NoStop}%
\bibitem [{\citenamefont {Hu}\ \emph {et~al.}(2022)\citenamefont {Hu} \emph {et~al.}}]{Hu:2021trw}%
  \BibitemOpen
  \bibfield  {author} {\bibinfo {author} {\bibfnamefont {B.}~\bibnamefont {Hu}} \emph {et~al.},\ }\bibfield  {title} {\bibinfo {title} {{Ab initio predictions link the neutron skin of $^{208}$Pb to nuclear forces}},\ }\href {https://doi.org/10.1038/s41567-023-02324-9} {\bibfield  {journal} {\bibinfo  {journal} {Nature Phys.}\ }\textbf {\bibinfo {volume} {18}},\ \bibinfo {pages} {1196} (\bibinfo {year} {2022})},\ \Eprint {https://arxiv.org/abs/2112.01125} {arXiv:2112.01125 [nucl-th]} \BibitemShut {NoStop}%
\bibitem [{\citenamefont {Dobaczewski}\ \emph {et~al.}(1984)\citenamefont {Dobaczewski}, \citenamefont {Flocard},\ and\ \citenamefont {Treiner}}]{Dobaczewski:1983zc}%
  \BibitemOpen
  \bibfield  {author} {\bibinfo {author} {\bibfnamefont {J.}~\bibnamefont {Dobaczewski}}, \bibinfo {author} {\bibfnamefont {H.}~\bibnamefont {Flocard}},\ and\ \bibinfo {author} {\bibfnamefont {J.}~\bibnamefont {Treiner}},\ }\bibfield  {title} {\bibinfo {title} {{Hartree-Fock-Bogolyubov descriptions of nuclei near the neutrino dripline}},\ }\href {https://doi.org/10.1016/0375-9474(84)90433-0} {\bibfield  {journal} {\bibinfo  {journal} {Nucl. Phys.}\ }\textbf {\bibinfo {volume} {A422}},\ \bibinfo {pages} {103} (\bibinfo {year} {1984})}\BibitemShut {NoStop}%
\bibitem [{\citenamefont {Bartel}\ \emph {et~al.}(1982)\citenamefont {Bartel}, \citenamefont {Quentin}, \citenamefont {Brack}, \citenamefont {Guet},\ and\ \citenamefont {Hakansson}}]{Bartel:1982ed}%
  \BibitemOpen
  \bibfield  {author} {\bibinfo {author} {\bibfnamefont {J.}~\bibnamefont {Bartel}}, \bibinfo {author} {\bibfnamefont {P.}~\bibnamefont {Quentin}}, \bibinfo {author} {\bibfnamefont {M.}~\bibnamefont {Brack}}, \bibinfo {author} {\bibfnamefont {C.}~\bibnamefont {Guet}},\ and\ \bibinfo {author} {\bibfnamefont {H.~B.}\ \bibnamefont {Hakansson}},\ }\bibfield  {title} {\bibinfo {title} {{Towards a better parametrisation of Skyrme-like effective forces: A Critical study of the SkM force}},\ }\href {https://doi.org/10.1016/0375-9474(82)90403-1} {\bibfield  {journal} {\bibinfo  {journal} {Nucl. Phys.}\ }\textbf {\bibinfo {volume} {A386}},\ \bibinfo {pages} {79} (\bibinfo {year} {1982})}\BibitemShut {NoStop}%
\bibitem [{\citenamefont {Kortelainen}\ \emph {et~al.}(2012)\citenamefont {Kortelainen}, \citenamefont {McDonnell}, \citenamefont {Nazarewicz}, \citenamefont {Reinhard}, \citenamefont {Sarich}, \citenamefont {Schunck}, \citenamefont {Stoitsov},\ and\ \citenamefont {Wild}}]{Kortelainen:2011ft}%
  \BibitemOpen
  \bibfield  {author} {\bibinfo {author} {\bibfnamefont {M.}~\bibnamefont {Kortelainen}}, \bibinfo {author} {\bibfnamefont {J.}~\bibnamefont {McDonnell}}, \bibinfo {author} {\bibfnamefont {W.}~\bibnamefont {Nazarewicz}}, \bibinfo {author} {\bibfnamefont {P.~G.}\ \bibnamefont {Reinhard}}, \bibinfo {author} {\bibfnamefont {J.}~\bibnamefont {Sarich}}, \bibinfo {author} {\bibfnamefont {N.}~\bibnamefont {Schunck}}, \bibinfo {author} {\bibfnamefont {M.~V.}\ \bibnamefont {Stoitsov}},\ and\ \bibinfo {author} {\bibfnamefont {S.~M.}\ \bibnamefont {Wild}},\ }\bibfield  {title} {\bibinfo {title} {{Nuclear energy density optimization: Large deformations}},\ }\href {https://doi.org/10.1103/PhysRevC.85.024304} {\bibfield  {journal} {\bibinfo  {journal} {Phys. Rev.}\ }\textbf {\bibinfo {volume} {C85}},\ \bibinfo {pages} {024304} (\bibinfo {year} {2012})},\ \Eprint {https://arxiv.org/abs/arXiv:1111.4344} {arXiv:1111.4344 [nucl-th]} \BibitemShut {NoStop}%
\bibitem [{\citenamefont {Kortelainen}\ \emph {et~al.}(2010)\citenamefont {Kortelainen}, \citenamefont {Lesinski}, \citenamefont {More}, \citenamefont {Nazarewicz}, \citenamefont {Sarich}, \citenamefont {Schunck}, \citenamefont {Stoitsov},\ and\ \citenamefont {Wild}}]{Kortelainen:2010hv}%
  \BibitemOpen
  \bibfield  {author} {\bibinfo {author} {\bibfnamefont {M.}~\bibnamefont {Kortelainen}}, \bibinfo {author} {\bibfnamefont {T.}~\bibnamefont {Lesinski}}, \bibinfo {author} {\bibfnamefont {J.}~\bibnamefont {More}}, \bibinfo {author} {\bibfnamefont {W.}~\bibnamefont {Nazarewicz}}, \bibinfo {author} {\bibfnamefont {J.}~\bibnamefont {Sarich}}, \bibinfo {author} {\bibfnamefont {N.}~\bibnamefont {Schunck}}, \bibinfo {author} {\bibfnamefont {M.~V.}\ \bibnamefont {Stoitsov}},\ and\ \bibinfo {author} {\bibfnamefont {S.}~\bibnamefont {Wild}},\ }\bibfield  {title} {\bibinfo {title} {{Nuclear Energy Density Optimization}},\ }\href {https://doi.org/10.1103/PhysRevC.82.024313} {\bibfield  {journal} {\bibinfo  {journal} {Phys. Rev.}\ }\textbf {\bibinfo {volume} {C82}},\ \bibinfo {pages} {024313} (\bibinfo {year} {2010})},\ \Eprint {https://arxiv.org/abs/arXiv:1005.5145} {arXiv:1005.5145 [nucl-th]} \BibitemShut {NoStop}%
\bibitem [{\citenamefont {Chabanat}\ \emph {et~al.}(1998)\citenamefont {Chabanat}, \citenamefont {Bonche}, \citenamefont {Haensel}, \citenamefont {Meyer},\ and\ \citenamefont {Schaeffer}}]{Chabanat:1997un}%
  \BibitemOpen
  \bibfield  {author} {\bibinfo {author} {\bibfnamefont {E.}~\bibnamefont {Chabanat}}, \bibinfo {author} {\bibfnamefont {P.}~\bibnamefont {Bonche}}, \bibinfo {author} {\bibfnamefont {P.}~\bibnamefont {Haensel}}, \bibinfo {author} {\bibfnamefont {J.}~\bibnamefont {Meyer}},\ and\ \bibinfo {author} {\bibfnamefont {R.}~\bibnamefont {Schaeffer}},\ }\bibfield  {title} {\bibinfo {title} {{A Skyrme parametrization from subnuclear to neutron star densities. 2. Nuclei far from stablities}},\ }\href {https://doi.org/10.1016/S0375-9474(98)00570-3, 10.1016/S0375-9474(98)00180-8} {\bibfield  {journal} {\bibinfo  {journal} {Nucl. Phys.}\ }\textbf {\bibinfo {volume} {A635}},\ \bibinfo {pages} {231} (\bibinfo {year} {1998})}\BibitemShut {NoStop}%
\bibitem [{\citenamefont {Reinhard}\ and\ \citenamefont {Flocard}(1995)}]{Reinhard:1995zz}%
  \BibitemOpen
  \bibfield  {author} {\bibinfo {author} {\bibfnamefont {P.~G.}\ \bibnamefont {Reinhard}}\ and\ \bibinfo {author} {\bibfnamefont {H.}~\bibnamefont {Flocard}},\ }\bibfield  {title} {\bibinfo {title} {{Nuclear effective forces and isotope shifts}},\ }\href {https://doi.org/10.1016/0375-9474(94)00770-N} {\bibfield  {journal} {\bibinfo  {journal} {Nucl. Phys.}\ }\textbf {\bibinfo {volume} {A584}},\ \bibinfo {pages} {467} (\bibinfo {year} {1995})}\BibitemShut {NoStop}%
\bibitem [{\citenamefont {Sharma}\ \emph {et~al.}(1993)\citenamefont {Sharma}, \citenamefont {Nagarajan},\ and\ \citenamefont {Ring}}]{Sharma:1993it}%
  \BibitemOpen
  \bibfield  {author} {\bibinfo {author} {\bibfnamefont {M.~M.}\ \bibnamefont {Sharma}}, \bibinfo {author} {\bibfnamefont {M.~A.}\ \bibnamefont {Nagarajan}},\ and\ \bibinfo {author} {\bibfnamefont {P.}~\bibnamefont {Ring}},\ }\bibfield  {title} {\bibinfo {title} {{rho meson coupling in the relativistic mean field theory and description of exotic nuclei}},\ }\href {https://doi.org/10.1016/0370-2693(93)90970-S} {\bibfield  {journal} {\bibinfo  {journal} {Phys. Lett.}\ }\textbf {\bibinfo {volume} {B312}},\ \bibinfo {pages} {377} (\bibinfo {year} {1993})}\BibitemShut {NoStop}%
\bibitem [{\citenamefont {Bender}\ \emph {et~al.}(1999)\citenamefont {Bender}, \citenamefont {Rutz}, \citenamefont {Reinhard}, \citenamefont {Maruhn},\ and\ \citenamefont {Greiner}}]{Bender:1999yt}%
  \BibitemOpen
  \bibfield  {author} {\bibinfo {author} {\bibfnamefont {M.}~\bibnamefont {Bender}}, \bibinfo {author} {\bibfnamefont {K.}~\bibnamefont {Rutz}}, \bibinfo {author} {\bibfnamefont {P.~G.}\ \bibnamefont {Reinhard}}, \bibinfo {author} {\bibfnamefont {J.~A.}\ \bibnamefont {Maruhn}},\ and\ \bibinfo {author} {\bibfnamefont {W.}~\bibnamefont {Greiner}},\ }\bibfield  {title} {\bibinfo {title} {{Shell structure of superheavy nuclei in selfconsistent mean field models}},\ }\href {https://doi.org/10.1103/PhysRevC.60.034304} {\bibfield  {journal} {\bibinfo  {journal} {Phys. Rev.}\ }\textbf {\bibinfo {volume} {C60}},\ \bibinfo {pages} {034304} (\bibinfo {year} {1999})},\ \Eprint {https://arxiv.org/abs/nucl-th/9906030} {nucl-th/9906030 [nucl-th]} \BibitemShut {NoStop}%
\bibitem [{\citenamefont {Lalazissis}\ \emph {et~al.}(1997)\citenamefont {Lalazissis}, \citenamefont {Konig},\ and\ \citenamefont {Ring}}]{Lalazissis:1996rd}%
  \BibitemOpen
  \bibfield  {author} {\bibinfo {author} {\bibfnamefont {G.~A.}\ \bibnamefont {Lalazissis}}, \bibinfo {author} {\bibfnamefont {J.}~\bibnamefont {Konig}},\ and\ \bibinfo {author} {\bibfnamefont {P.}~\bibnamefont {Ring}},\ }\bibfield  {title} {\bibinfo {title} {{A New parametrization for the Lagrangian density of relativistic mean field theory}},\ }\href {https://doi.org/10.1103/PhysRevC.55.540} {\bibfield  {journal} {\bibinfo  {journal} {Phys. Rev.}\ }\textbf {\bibinfo {volume} {C55}},\ \bibinfo {pages} {540} (\bibinfo {year} {1997})},\ \Eprint {https://arxiv.org/abs/nucl-th/9607039} {nucl-th/9607039 [nucl-th]} \BibitemShut {NoStop}%
\bibitem [{\citenamefont {Reinhard}\ \emph {et~al.}(1986)\citenamefont {Reinhard}, \citenamefont {Rufa}, \citenamefont {Maruhn}, \citenamefont {Greiner},\ and\ \citenamefont {Friedrich}}]{Reinhard:1986qq}%
  \BibitemOpen
  \bibfield  {author} {\bibinfo {author} {\bibfnamefont {P.~G.}\ \bibnamefont {Reinhard}}, \bibinfo {author} {\bibfnamefont {M.}~\bibnamefont {Rufa}}, \bibinfo {author} {\bibfnamefont {J.}~\bibnamefont {Maruhn}}, \bibinfo {author} {\bibfnamefont {W.}~\bibnamefont {Greiner}},\ and\ \bibinfo {author} {\bibfnamefont {J.}~\bibnamefont {Friedrich}},\ }\bibfield  {title} {\bibinfo {title} {{Nuclear Ground State Properties in a Relativistic Meson Field Theory}},\ }\href@noop {} {\bibfield  {journal} {\bibinfo  {journal} {Z. Phys. A}\ }\textbf {\bibinfo {volume} {323}},\ \bibinfo {pages} {13} (\bibinfo {year} {1986})}\BibitemShut {NoStop}%
\bibitem [{\citenamefont {Niksic}\ \emph {et~al.}(2008)\citenamefont {Niksic}, \citenamefont {Vretenar},\ and\ \citenamefont {Ring}}]{Niksic:2008vp}%
  \BibitemOpen
  \bibfield  {author} {\bibinfo {author} {\bibfnamefont {T.}~\bibnamefont {Niksic}}, \bibinfo {author} {\bibfnamefont {D.}~\bibnamefont {Vretenar}},\ and\ \bibinfo {author} {\bibfnamefont {P.}~\bibnamefont {Ring}},\ }\bibfield  {title} {\bibinfo {title} {{Relativistic Nuclear Energy Density Functionals: Adjusting parameters to binding energies}},\ }\href {https://doi.org/10.1103/PhysRevC.78.034318} {\bibfield  {journal} {\bibinfo  {journal} {Phys. Rev.}\ }\textbf {\bibinfo {volume} {C78}},\ \bibinfo {pages} {034318} (\bibinfo {year} {2008})},\ \Eprint {https://arxiv.org/abs/arXiv:0809.1375} {arXiv:0809.1375 [nucl-th]} \BibitemShut {NoStop}%
\bibitem [{\citenamefont {Niksic}\ \emph {et~al.}(2002)\citenamefont {Niksic}, \citenamefont {Vretenar}, \citenamefont {Finelli},\ and\ \citenamefont {Ring}}]{Niksic:2002yp}%
  \BibitemOpen
  \bibfield  {author} {\bibinfo {author} {\bibfnamefont {T.}~\bibnamefont {Niksic}}, \bibinfo {author} {\bibfnamefont {D.}~\bibnamefont {Vretenar}}, \bibinfo {author} {\bibfnamefont {P.}~\bibnamefont {Finelli}},\ and\ \bibinfo {author} {\bibfnamefont {P.}~\bibnamefont {Ring}},\ }\bibfield  {title} {\bibinfo {title} {{Relativistic Hartree-Bogolyubov model with density dependent meson nucleon couplings}},\ }\href {https://doi.org/10.1103/PhysRevC.66.024306} {\bibfield  {journal} {\bibinfo  {journal} {Phys. Rev.}\ }\textbf {\bibinfo {volume} {C66}},\ \bibinfo {pages} {024306} (\bibinfo {year} {2002})},\ \Eprint {https://arxiv.org/abs/nucl-th/0205009} {nucl-th/0205009 [nucl-th]} \BibitemShut {NoStop}%
\bibitem [{\citenamefont {Hernandez}(2019)}]{Hernandez:2019hsk}%
  \BibitemOpen
  \bibfield  {author} {\bibinfo {author} {\bibfnamefont {J.~A.}\ \bibnamefont {Hernandez}},\ }\href@noop {} {\bibinfo {title} {{Weak Nuclear Form Factor: Nuclear Structure and Coherent Elastic Neutrino-Nucleus Scattering, Master's Thesis}}} (\bibinfo {year} {2019})\BibitemShut {NoStop}%
\bibitem [{\citenamefont {Yang}\ \emph {et~al.}(2019)\citenamefont {Yang}, \citenamefont {Hernandez},\ and\ \citenamefont {Piekarewicz}}]{Yang:2019pbx}%
  \BibitemOpen
  \bibfield  {author} {\bibinfo {author} {\bibfnamefont {J.}~\bibnamefont {Yang}}, \bibinfo {author} {\bibfnamefont {J.~A.}\ \bibnamefont {Hernandez}},\ and\ \bibinfo {author} {\bibfnamefont {J.}~\bibnamefont {Piekarewicz}},\ }\bibfield  {title} {\bibinfo {title} {{Electroweak probes of ground state densities}},\ }\href {https://doi.org/10.1103/PhysRevC.100.054301} {\bibfield  {journal} {\bibinfo  {journal} {Phys. Rev. C}\ }\textbf {\bibinfo {volume} {100}},\ \bibinfo {pages} {054301} (\bibinfo {year} {2019})},\ \Eprint {https://arxiv.org/abs/1908.10939} {arXiv:1908.10939 [nucl-th]} \BibitemShut {NoStop}%
\bibitem [{\citenamefont {Chen}\ and\ \citenamefont {Piekarewicz}(2014)}]{Chen:2014sca}%
  \BibitemOpen
  \bibfield  {author} {\bibinfo {author} {\bibfnamefont {W.-C.}\ \bibnamefont {Chen}}\ and\ \bibinfo {author} {\bibfnamefont {J.}~\bibnamefont {Piekarewicz}},\ }\bibfield  {title} {\bibinfo {title} {{Building relativistic mean field models for finite nuclei and neutron stars}},\ }\href {https://doi.org/10.1103/PhysRevC.90.044305} {\bibfield  {journal} {\bibinfo  {journal} {Phys. Rev. C}\ }\textbf {\bibinfo {volume} {90}},\ \bibinfo {pages} {044305} (\bibinfo {year} {2014})},\ \Eprint {https://arxiv.org/abs/1408.4159} {arXiv:1408.4159 [nucl-th]} \BibitemShut {NoStop}%
\bibitem [{\citenamefont {Chen}\ and\ \citenamefont {Piekarewicz}(2015)}]{Chen:2014mza}%
  \BibitemOpen
  \bibfield  {author} {\bibinfo {author} {\bibfnamefont {W.-C.}\ \bibnamefont {Chen}}\ and\ \bibinfo {author} {\bibfnamefont {J.}~\bibnamefont {Piekarewicz}},\ }\bibfield  {title} {\bibinfo {title} {{Searching for isovector signatures in the neutron-rich oxygen and calcium isotopes}},\ }\href {https://doi.org/10.1016/j.physletb.2015.07.020} {\bibfield  {journal} {\bibinfo  {journal} {Phys. Lett. B}\ }\textbf {\bibinfo {volume} {748}},\ \bibinfo {pages} {284} (\bibinfo {year} {2015})},\ \Eprint {https://arxiv.org/abs/1412.7870} {arXiv:1412.7870 [nucl-th]} \BibitemShut {NoStop}%
\bibitem [{\citenamefont {Zheng}\ \emph {et~al.}(2014)\citenamefont {Zheng}, \citenamefont {Zhang},\ and\ \citenamefont {Chen}}]{Zheng:2014nga}%
  \BibitemOpen
  \bibfield  {author} {\bibinfo {author} {\bibfnamefont {H.}~\bibnamefont {Zheng}}, \bibinfo {author} {\bibfnamefont {Z.}~\bibnamefont {Zhang}},\ and\ \bibinfo {author} {\bibfnamefont {L.-W.}\ \bibnamefont {Chen}},\ }\bibfield  {title} {\bibinfo {title} {{Form Factor Effects in the Direct Detection of Isospin-Violating Dark Matter}},\ }\href {https://doi.org/10.1088/1475-7516/2014/08/011} {\bibfield  {journal} {\bibinfo  {journal} {JCAP}\ }\textbf {\bibinfo {volume} {08}},\ \bibinfo {pages} {011}},\ \Eprint {https://arxiv.org/abs/1403.5134} {arXiv:1403.5134 [nucl-th]} \BibitemShut {NoStop}%
\bibitem [{\citenamefont {Sil}\ \emph {et~al.}(2005)\citenamefont {Sil}, \citenamefont {Centelles}, \citenamefont {Vinas},\ and\ \citenamefont {Piekarewicz}}]{Sil:2005tg}%
  \BibitemOpen
  \bibfield  {author} {\bibinfo {author} {\bibfnamefont {T.}~\bibnamefont {Sil}}, \bibinfo {author} {\bibfnamefont {M.}~\bibnamefont {Centelles}}, \bibinfo {author} {\bibfnamefont {X.}~\bibnamefont {Vinas}},\ and\ \bibinfo {author} {\bibfnamefont {J.}~\bibnamefont {Piekarewicz}},\ }\bibfield  {title} {\bibinfo {title} {{Atomic parity non-conservation, neutron radii, and effective field theories of nuclei}},\ }\href {https://doi.org/10.1103/PhysRevC.71.045502} {\bibfield  {journal} {\bibinfo  {journal} {Phys. Rev. C}\ }\textbf {\bibinfo {volume} {71}},\ \bibinfo {pages} {045502} (\bibinfo {year} {2005})},\ \Eprint {https://arxiv.org/abs/nucl-th/0501014} {arXiv:nucl-th/0501014} \BibitemShut {NoStop}%
\bibitem [{\citenamefont {Yue}\ \emph {et~al.}(2022)\citenamefont {Yue}, \citenamefont {Chen}, \citenamefont {Zhang},\ and\ \citenamefont {Zhou}}]{Yue:2021yfx}%
  \BibitemOpen
  \bibfield  {author} {\bibinfo {author} {\bibfnamefont {T.-G.}\ \bibnamefont {Yue}}, \bibinfo {author} {\bibfnamefont {L.-W.}\ \bibnamefont {Chen}}, \bibinfo {author} {\bibfnamefont {Z.}~\bibnamefont {Zhang}},\ and\ \bibinfo {author} {\bibfnamefont {Y.}~\bibnamefont {Zhou}},\ }\bibfield  {title} {\bibinfo {title} {{Constraints on the symmetry energy from PREX-II in the multimessenger era}},\ }\href {https://doi.org/10.1103/PhysRevResearch.4.L022054} {\bibfield  {journal} {\bibinfo  {journal} {Phys. Rev. Res.}\ }\textbf {\bibinfo {volume} {4}},\ \bibinfo {pages} {L022054} (\bibinfo {year} {2022})},\ \Eprint {https://arxiv.org/abs/2102.05267} {arXiv:2102.05267 [nucl-th]} \BibitemShut {NoStop}%
\end{thebibliography}%

\end{document}